\documentclass[twocolumn]{aastex701}
\usepackage{lipsum}
\usepackage{graphicx}
\usepackage{subcaption}
\usepackage{amsmath}
\usepackage{CJK}

\begin{document}

\title{AT~2024ahzi: A Type IIP Supernova Discovered by the LSST Commissioning Camera}

\author[orcid=0000-0002-9886-2834,sname='de~Soto']{Kaylee~de~Soto}
\affiliation{Center for Astrophysics \textbar{} Harvard \& Smithsonian, 60 Garden Street, Cambridge, MA 02138-1516, USA}
\email[show]{kaylee.de\_soto@cfa.harvard.edu} 

\author[orcid=0000-0002-5814-4061, sname='Villar', gname='Victoria Ashley']{V.~Ashley~Villar}
\affiliation{Center for Astrophysics \textbar{} Harvard \& Smithsonian, 60 Garden Street, Cambridge, MA 02138-1516, USA}
\affiliation{The NSF AI Institute for Artificial Intelligence and Fundamental Interactions}
\email{ashleyvillar@cfa.harvard.edu} 

\author[orcid=0000-0003-1012-3031,sname='Goldberg']{Jared~A.~Goldberg}
\affil{Department of Physics and Astronomy, Michigan State University, East Lansing, MI 48824, USA}
\affil{Center for Computational Astrophysics, Flatiron Institute, 162 5th Avenue, New York, NY 10010, USA}
\email{goldstar@msu.edu}

\author[orcid=0000-0002-2028-9329,sname='Nugent']{Anya~Nugent}
\affiliation{Center for Astrophysics \textbar{} Harvard \& Smithsonian, 60 Garden Street, Cambridge, MA 02138-1516, USA}
\email{anya.nugent@cfa.harvard.edu}

\begin{CJK*}{UTF8}{gbsn}
\author[0000-0002-7937-6371,sname=Yize,gname=Dong]{Yize Dong (董一泽)}
\affiliation{Center for Astrophysics \textbar{} Harvard \& Smithsonian, 60 Garden Street, Cambridge, MA 02138-1516, USA}
\affiliation{The NSF AI Institute for Artificial Intelligence and Fundamental Interactions}
\email{yize.dong@cfa.harvard.edu}

\author[orcid=0000-0002-2445-5275,sname='Foley']{Ryan~J.~Foley}
\affiliation{Department of Astronomy and Astrophysics, University of California, Santa Cruz, CA 95064, USA}
\email{foley@ucsc.edu}

\author[orcid=0000-0002-6851-9613, sname="G\'eron"]{Tobias~G\'eron}
\affiliation{David A. Dunlap Institute of Astronomy \& Astrophysics
University of Toronto, Toronto, ON, M5S 3H4, Canada}
\email{tobias.geron@utoronto.ca}

\author[orcid=0000-0001-9695-8472,sname='Izzo']{Luca~Izzo}
\affiliation{DARK, Niels Bohr Institute, University of Copenhagen, Jagtvej 128, 2200 Copenhagen, Denmark}
\affiliation{INAF, Osservatorio Astronomico di Capodimonte, salita Moiariello 16, I-80131, Naples, Italy}
\email{luca.izzo@inaf.it}

\author[0009-0006-5214-0736]{C.~Tanner~Murphey}
\affiliation{Department of Astronomy, University of Illinois at Urbana-Champaign, 1002 W. Green St., IL 61801, USA}
\affiliation{Center for Astrophysical Surveys, National Center for Supercomputing Applications, Urbana, IL, 61801, USA}
\affiliation{Illinois Center for Advanced Studies of the Universe, Urbana, IL 61801}
\email{murphey2@illinois.edu}

\author[orcid=0000-0002-4449-9152,sname='Auchettl']{Katie~Auchettl}
\affiliation{Department of Astronomy and Astrophysics, University of California, Santa Cruz, CA 95064, USA}
\affiliation{OzGrav, School of Physics, The University of Melbourne, VIC 3010, Australia}
\email{katie.auchettl@unimelb.edu.au}

\author[orcid=0000-0003-4263-2228,sname='Coulter']{David~A.~Coulter}
\affiliation{William H. Miller III Department of Physics \& Astronomy, Johns Hopkins University, 3400 N Charles St, Baltimore, MD 21218, USA}
\affiliation{Space Telescope Science Institute, Baltimore, MD 21218, USA}
\email{dcoulter@stsci.edu}

\author[orcid=0000-0001-5486-2747,sname='de Boer']{Thomas~de~Boer}
\affiliation{Institute for Astronomy, University of Hawai'i, 2680 Woodlawn Drive, Honolulu, HI 96822, USA}
\email{tdeboer@hawaii.edu}

\author[orcid=0000-0001-6965-7789,sname='Chambers']{Kenneth~C.~Chambers}
\affiliation{Institute for Astronomy, University of Hawai'i, 2680 Woodlawn Drive, Honolulu, HI 96822, USA}
\email{kchamber@hawaii.edu}

\author[orcid=0000-0002-6886-269X,sname='Farias']{Diego~A.~Farias}
\affiliation{DARK, Niels Bohr Institute, University of Copenhagen, Jagtvej 128, 2200 Copenhagen, Denmark}
\email{diego.farias@nbi.ku.dk}

\author[orcid=0000-0002-8526-3963,sname='Gall']{Christa~Gall}
\affiliation{DARK, Niels Bohr Institute, University of Copenhagen, Jagtvej 128, 2200 Copenhagen, Denmark}
\email{christa.gall@nbi.ku.dk}

\author[orcid=0000-0003-1015-5367,sname='Gao']{Hua~Gao}
\affiliation{Institute for Astronomy, University of Hawai'i, 2680 Woodlawn Drive, Honolulu, HI 96822, USA}
\email{hgao@hawaii.edu}

\author[orcid=0000-0002-4571-2306,sname='Hjorth']{Jens~Hjorth}
\affiliation{DARK, Niels Bohr Institute, University of Copenhagen, Jagtvej 128, 2200 Copenhagen, Denmark}
\email{jens@nbi.ku.dk}

\author[orcid=0000-0003-3953-9532,sname='Hoogendam']{Willem~B.~Hoogendam}
\affiliation{Institute for Astronomy, University of Hawai'i, 2680 Woodlawn Drive, Honolulu, HI 96822, USA}
\email{willemh@hawaii.edu}

\author[orcid=0000-0002-6230-0151,sname='Jones']{David~O.~Jones}
\affiliation{Institute for Astronomy, University of Hawai'i, 640 N.~Aʻohoku Pl., Hilo, HI 96720, USA}
\email{dojones@hawaii.edu}

\author[0009-0009-9760-0718,sname='Nair']{Gauri~Nair}
\affiliation{Department of Astronomy, University of Illinois at Urbana-Champaign, 1002 W. Green St., IL 61801, USA}
\email{gaurin2@illinois.edu}

\author[orcid=0000-0001-6022-0484,sname='Narayan']{Gautham~Narayan}
\affiliation{Department of Astronomy, University of Illinois at Urbana-Champaign, 1002 W. Green St., IL 61801, USA}
\affiliation{Center for Astrophysical Surveys, National Center for Supercomputing Applications, Urbana, IL, 61801, USA}
\affiliation{NSF-Simons SkAI Institute, 875 N. Michigan Ave., Chicago, IL 60611, USA}
\email{gsn@illinois.edu}

\author[orcid=0000-0002-4410-5387,sname='Rest']{Armin~Rest}
\affiliation{William H. Miller III Department of Physics \& Astronomy, Johns Hopkins University, 3400 N Charles St, Baltimore, MD 21218, USA}
\affiliation{Space Telescope Science Institute, Baltimore, MD 21218, USA}
\email{arest@stsci.edu}

\author[orcid=0000-0002-1092-6806,sname='Patra']{Kishore~C.~Patra}
\affiliation{Department of Astronomy and Astrophysics, University of California, Santa Cruz, CA 95064, USA}
\email{kcpatra@ucsc.edu}

\author[orcid=0009-0000-5561-9116,sname='Perkins']{Haille~M.~L.~Perkins}
\affiliation{Department of Astronomy, University of Illinois at Urbana-Champaign, 1002 W. Green St., IL 61801, USA}
\email{haillep2@illinois.edu}

\author[orcid=0000-0003-1535-4277,sname='Verrico']{Margaret~E.~Verrico}
\affiliation{Department of Astronomy, University of Illinois at Urbana-Champaign, 1002 W. Green St., IL 61801, USA}
\email{verrico2@illinois.edu}

\author[orcid=0000-0001-5233-6989,sname='Wang']{Qinan~Wang}
\affiliation{Department of Physics and Kavli Institute for Astrophysics and Space Research, Massachusetts Institute of Technology, 77 Massachusetts Avenue, Cambridge, MA 02139, USA}
\email{qnwang@mit.edu}

\author[orcid=0000-0002-4186-6164,sname='Wasserman']{Amanda~R.~Wasserman}
\affiliation{Department of Astronomy, University of Illinois at Urbana-Champaign, 1002 W. Green St., IL 61801, USA}
\email{amandaw8@illinois.edu}

\author[orcid=0000-0002-0632-8897,sname='Zenati']{Yossef~Zenati}
\affiliation{Astrophysics Research Center of the Open University (ARCO), Department of Natural Sciences, Ra’anana 4353701, Israel}
\affiliation{William H. Miller III Department of Physics \& Astronomy, Johns Hopkins University, 3400 N Charles St, Baltimore, MD 21218, USA}
\email{yzenati1@jhu.edu}

\begin{abstract}
As part of its commissioning, the Vera C.\ Rubin Observatory observed several fields repeatedly for a month with ComCam, an instrument that uses the same hardware as the LSST camera but covers a smaller field of view. We photometrically classify AT~2024ahzi, a transient discovered by ComCam, as a Type IIP supernova (SN~IIP) using both ComCam and DECam photometry. We find that the duration, luminosity, and color of AT~2024ahzi's photometric plateau are all consistent with those from a large sample of SNe~II. By comparing its multi-band light curves to SN~II models and analytic relations, we place constraints on the SN progenitor, explosion dynamics, and circumstellar environment. We argue that the progenitor has an extended density profile indistinguishable from a slowly accelerating CSM. We discuss how a similar workflow can identify and characterize future Rubin SNe~II.
\end{abstract}

\keywords{}


\section{Introduction}

Core-collapse supernovae (CCSNe) are the endpoints of massive stars and are important drivers of the chemical evolution of the Cosmos \citep{smartt_2009, jerkstrand2025}. The most common subclass of CCSNe, SNe~IIP, are characterized by months-long plateaus in their light curves from the cooling and recombination of hydrogen-rich envelopes \citep{filippenko_1997, smartt_2009, li_2011, smith_2011, galyam_2017}. SNe~IIP are traditionally contrasted with SNe~IIL, which have a linear brightness decline and tend to be brighter than SNe~IIP \citep{barbon_1979, patat_1994}. Recent studies suggest that SNe~IIP and SNe~IIL form a continuum rather than two distinct subclasses \citep{Anderson_2014, gall_2015, Sanders_2015, Galbany_2016, valenti2016, Anderson_2024} The brightness, duration, and multiband color of SN~IIP plateaus provide insight into their underlying explosion dynamics and progenitor properties \citep{popov_1993, chieffi_2003, hamuy_2003, Young_2004, Pejcha_2015, goldberg_2019, Hiramatsu_2021, Martinez_csp_III}.

There is increasing evidence for dense circumstellar matter (CSM) surrounding CCSNe, formed from mass lost via stellar winds, stellar outbursts, or binary envelope stripping \citep{yaron_2017, Kozyreva_2022, irani_2024, jacobson_2024, iip_csm_ztf, nagao_2025}. The supernova ejecta's interaction with the CSM manifests as a bright blue continuum with or without narrow Lorentzian emission lines, as photons scatter off free electrons before leaving the CSM \citep{Chugai_2001}. The prevalence of CSM around SNe~II, and whether it explains the continuum between SNe~IIP and SNe~IIL, both remain open questions \citep{Morozova_2017, Morozova_2018, bostroem_2019, Hiramatsu_2021}. Early high-ionization flash features and rapid optical rise timescales have been found in $>$36\% \citep{Khazov_2016, Bruch_2021, bruch_2023} and $>90\%$ of typical SNe~II \citep{Forster2018}, respectively, both indicative of a confined CSM \citep{falk_1977, Nagy_2016, hillier_2019, Bruch_2021}.  

There have been $\sim$4,000 detected and spectroscopically classified SNe~II so far, with most detected by the Zwicky Transient Facility (ZTF; \citealt{ztf, Perley_2020}) and the Asteroid Terrestrial-impact Last Alert System (ATLAS; \citealt{atlas})\footnote{\url{www.wis-tns.org}}. Both observe the night sky every two days in $gr$- and cyan/orange bands, respectively, with ZTF limited to the northern sky. While two bands are often sufficient for classification, they cannot robustly constrain both blackbody temperature and photospheric radius. Along the hydrogen-recombination plateau (at $~6000$ K), the $g-$ and $r-$bands straddle the continuum peak, so extracted blackbody temperatures are more confident. At early phases, however, blackbody temperatures exceed $\sim$10,000 K, and both wavelengths lie on the Rayleigh-Jeans tail. This, along with biases from optical emission lines and UV iron-element line blanketing, means that early-phase blackbody temperatures are overestimated without near-UV observations. Previous works find biases of ${\sim}$10-50\% \citep{dessart_2005, valenti2016, Faran_2018}.

The Vera C.\ Rubin Observatory Legacy Survey of Space and Time (LSST) will detect SNe at a limiting flux $\sim$40 times fainter than ZTF \citep{lsst_science_book_2009}. It is expected to find $\sim$300,000 CCSNe annually \citep{Kessler_2019}, at a median redshift of $\sim$0.2. This two-order-of-magnitude increase in sample size will allow us to characterize the diversity in photometric evolution among SNe~IIP to high precision, especially at redshifts beyond 0.1. Establishing relations between progenitor properties and SN~II photometry will allow for more certain conclusions regarding the perceived lack of $> 18 M_\odot$ SN~II progenitors (dubbed the missing red supergiant problem; \citealt{smartt_2009, smartt_2015, Kochanek_2020}). While this discrepancy between RSG and SN II progenitor mass limits can be attributed to failed SNe \citep{Kochanek_2008, Adams_2017, Neustadt_2021}, it can also be attributed to observational biases \citep{walmswell_2012, davies_beasor_2018, Kilpatrick_2023, beasor_2024, strotjohann2024} or small-number statistics \citep{davies_beasor_2020, healy2024}. Large photometric SN~II samples could better populate the high-mass progenitor tail by assuming a correlation between plateau properties and progenitor mass \citep{kasen_2009, Dessart_2010, goldberg_2019, Barker_2022, Barker_2023}.

LSST will cycle between $ugrizy$ filters. At typical expected redshifts, the $u$-band will probe the near-to-mid UV, which aligns with temperatures of shock breakout augmented by CSM \citep{Waxman_2017, irani_2024}. The near-IR bands will improve blackbody fits by avoiding iron-element line blanketing \citep{Martinez_2022_a} and experiencing less extinction from dust.

As part of preparation for LSST, the Rubin commissioning camera (dubbed ComCam), conducted preliminary science observations from November 4, 2024 to December 10, 2024, collating all detections and forced photometry in the Data Preview 1 (DP1; \citealt{dp1, dp1_2}). These observations were taken in a handful of fields at a $\sim$1-3 day cadence.

In this paper, we photometrically classify and analyze AT~2024ahzi, making it the first photometrically classified SN~IIP after detection by the Rubin Observatory. We simultaneously observed AT~2024ahzi with the Dark Energy Camera (DECam), which captured its full plateau and increased classification confidence.

In Section~\ref{sec:data} we detail how we collated DECam photometry, prioritized SN candidates, and retrieved ComCam photometry for overlapping events. In Section~\ref{sec:host}, we associate AT~2024ahzi to a host galaxy and infer the host's redshift, dust extinction, and star formation history. In Section~\ref{sec:analysis}, we model AT~2024ahzi's joint ComCam-DECam photometry to characterize its evolution over time and estimate its circumstellar and progenitor properties. In Section~\ref{sec:pop_compare}, we show that AT~2024ahzi's plateau properties are congruent with those from a large, low-$z$ SN~IIP population. We summarize our findings in Section~\ref{sec:conc} and provide recommendations for future Rubin SN~IIP identification.

\section{Object Selection and Data Collation} \label{sec:data}

\subsection{Transient Identification with DECam}

The Young Supernova Experiment (YSE; \citealt{yse}) collaboration used the Dark Energy Camera (DECam; \citealt{DES2015}) mounted on the 4-m Victor M.\ Blanco Telescope at the Cerro Tololo Inter-American Observatory (CTIO) to observe the Euclid Deep Field South (EDFS) and Extended Chandra Deep Field South (ECDFS) fields (NOIRLab PropIDs 2024B-763968, 2024B-441839, and 2025A-388357). YSE observed the EDFS field 414 times in \textit{griz} between November 26, 2024 and April 27, 2025, and the ECDFS field 366 times between August 22, 2024 and April 18, 2025. Data were initially processed by the NOIRLab Community Pipeline and then transferred to the Illinois Campus Cluster. The images were then processed using the \texttt{Photpipe} difference imaging and forced photometry pipelines \citep{Rest2004,Rest2014}.

After performing standard data reductions on NOIRLab images (i.e., masking saturated pixels, correcting optical distortions, and basic photometry using \texttt{DoPHOT}; \citealt{dophot}), \texttt{Photpipe} subtracts template images from science images using \texttt{HOTPANTS} \citep{hotpants}. Template images are selected to avoid including flux from the transients themselves; identification of AT~2024ahzi used templates from November 6, 2024. Possible transients are identified as coordinates with at least three $3\sigma$ detections across multiple consecutive observing nights. YSE astronomers manually filter these candidates by examining the forced photometry and difference images.

AT~2024ahzi was first detected on December 19, 2024 in DECam's \textit{Euclid} Deep Field South (EDFS) field, identified by YSE astronomers on January 22, 2025, and reported to the \textit{Transient Name Server} (TNS) on March 13, 2025 \citep{tns-report}. It is located at R.A. = $03^{\text h} 53 ^{\text m} 20^{\text s}.41$, Dec = $-48^{\circ} 45' 01''.09$.

\subsection{Cross-matching with ComCam Objects}

\begin{figure*}[t]
    \centering
    \begin{subfigure}[t]{0.55\textwidth}
        \centering
    \includegraphics[width=\textwidth]{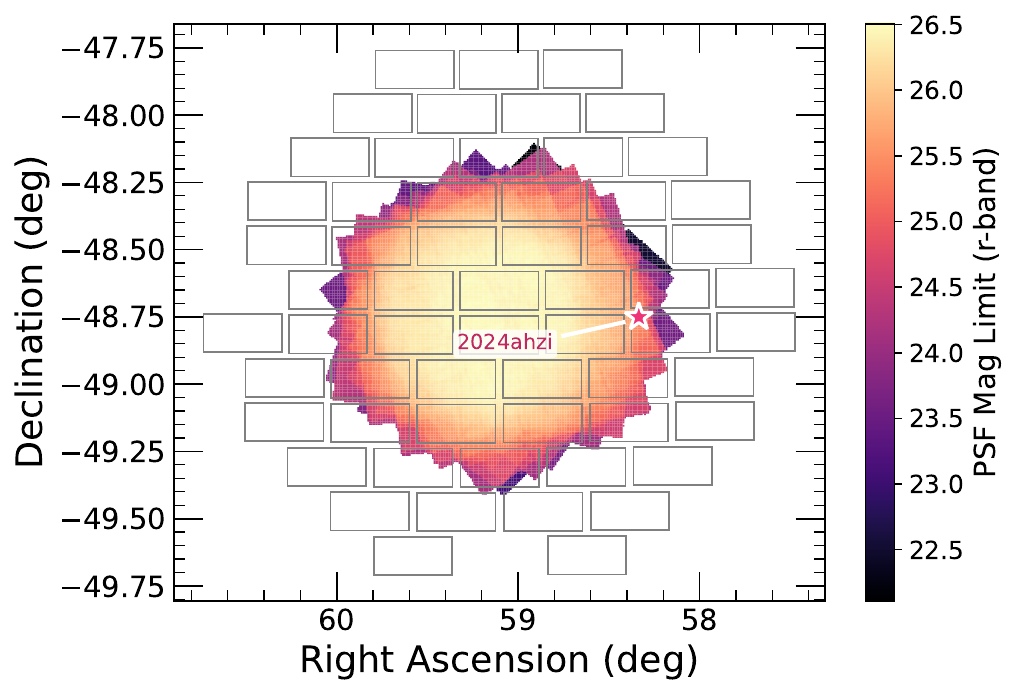}
        \label{fig:bottom_left}
    \end{subfigure}
    \begin{subfigure}[t]{0.4\textwidth}
        \centering
        \includegraphics[width=\textwidth]{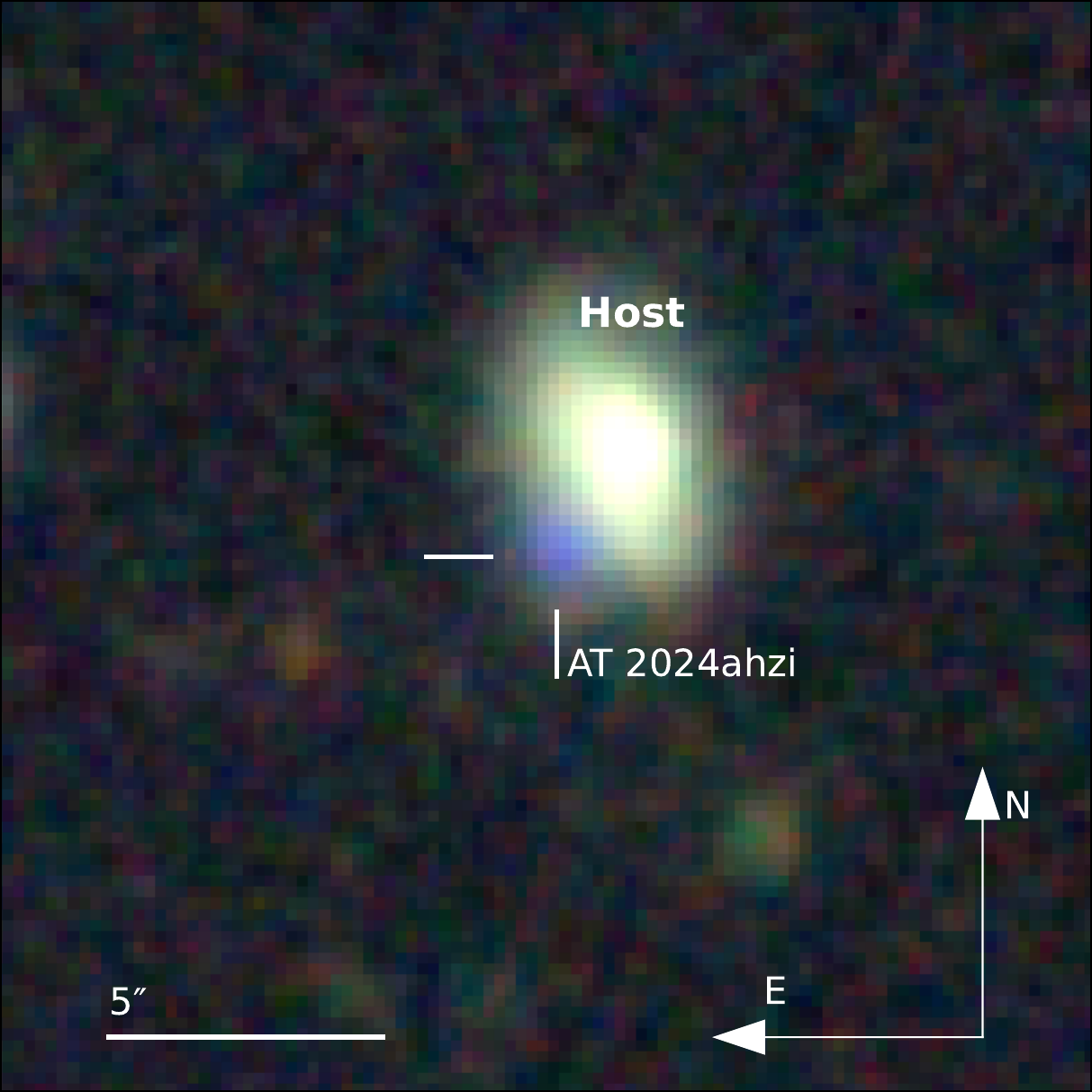}
        \label{fig:bottom_right}
    \end{subfigure}
    \caption{{\it Left}: ComCam exposure map showing the location of AT~2024ahzi and a single DECam pointing overlaid. Grey rectangles represent single DECam CCDs. {\it Right:} A ComCam $gri$ color-composite image of the field surrounding AT~2024ahzi. The SN is circled and the crosshairs show the location of the host, which has a spectroscopic redshift $z=0.211\pm0.002$.}
    \label{fig:host}
\end{figure*}

ComCam obtained $ugrizy$ observations of seven fields from November 4, 2024 to December 10, 2024 \citep{LSST_SITCOMTN149} as part of LSST commissioning. Concurrent DECam observations overlap with the ComCam ECDFS and EDFS fields; we show the EDFS overlap in Figure~\ref{fig:host}. We only consider DECam candidates within this overlapping region that were observed at least once between July 27, 2024 and January 9, 2025, as a generous buffer around the ComCam observing window. These constraints yield 61 DECam candidates in ECDFS and 8 in EDFS.

We then query the DP1 \texttt{DiaObject} catalog for all objects with difference-image detections that coincide with our remaining DECam candidates. We find matches for 7 events in ECDFS and 2 events in EDFS. Among these, AT~2024ahzi (designated by Rubin as LSST-DP1-DO-592914119179370575) shows clear SN~IIP-like evolution in the joint ComCam-DECam photometry. The remaining events are analyzed in \cite{tanner_in_prep}.

\subsection{Template Correction and Pre-processing}

\begin{deluxetable*}{l|c|c|c|c|c|c}
    \centering
    \caption{Subset of nightly co-added observations of AT~2024ahzi, uncorrected for Milky Way or host extinction. We include the first $5\sigma$ detection preceded by the last two non-detections. All uncertainties are $1\sigma$. The full single-exposure and nightly co-added photometry are available as machine readable tables online.}
    \label{tab:phottab}
    \tablehead{
        \colhead{MJD} & \colhead{Instrument} & \colhead{Filter} & \colhead{AB Mag} & \colhead{AB Mag Unc.} & \colhead{Flux ($\mu$Jy)} & \colhead{Flux Unc. ($\mu$Jy)}
    }
    \startdata
        60641.110 & ComCam & i & 24.188\tablenotemark{\dag} & --- & -0.307 & 0.153 \\
        60641.119 & ComCam & r & 25.888\tablenotemark{\dag} & --- & -0.366 & 0.032 \\
        60647.135 & ComCam & r & 22.664 & 0.088 & 3.121 & 0.253 \\
        60649.257 & DECam & g & 22.111 & 0.083 & 5.196 & 0.397 \\
        60649.258 & DECam & r & 22.413 & 0.195 & 3.935 & 0.708 \\
        60649.259 & DECam & i & 22.406 & 0.181 & 3.959 & 0.660 \\
        60652.170 & DECam & g & 22.353 & 0.111 & 4.159 & 0.426 \\
        60652.171 & DECam & r & 22.454 & 0.122 & 3.787 & 0.426 \\
        60652.172 & DECam & i & 22.367 & 0.167 & 4.103 & 0.632 \\
        60653.233 & ComCam & r & 22.430 & 0.054 & 3.874 & 0.191 \\
        60653.234 & ComCam & g & 22.223 & 0.032 & 4.685 & 0.137 \\
        60653.240 & ComCam & z & 22.460 & 0.148 & 3.768 & 0.514
    \enddata
    \tablenotetext{$\dag$}{5-$\sigma$ non-detection upper limit}
\end{deluxetable*}

\begin{figure*}[t]
    \centering
    \includegraphics[width=\textwidth]{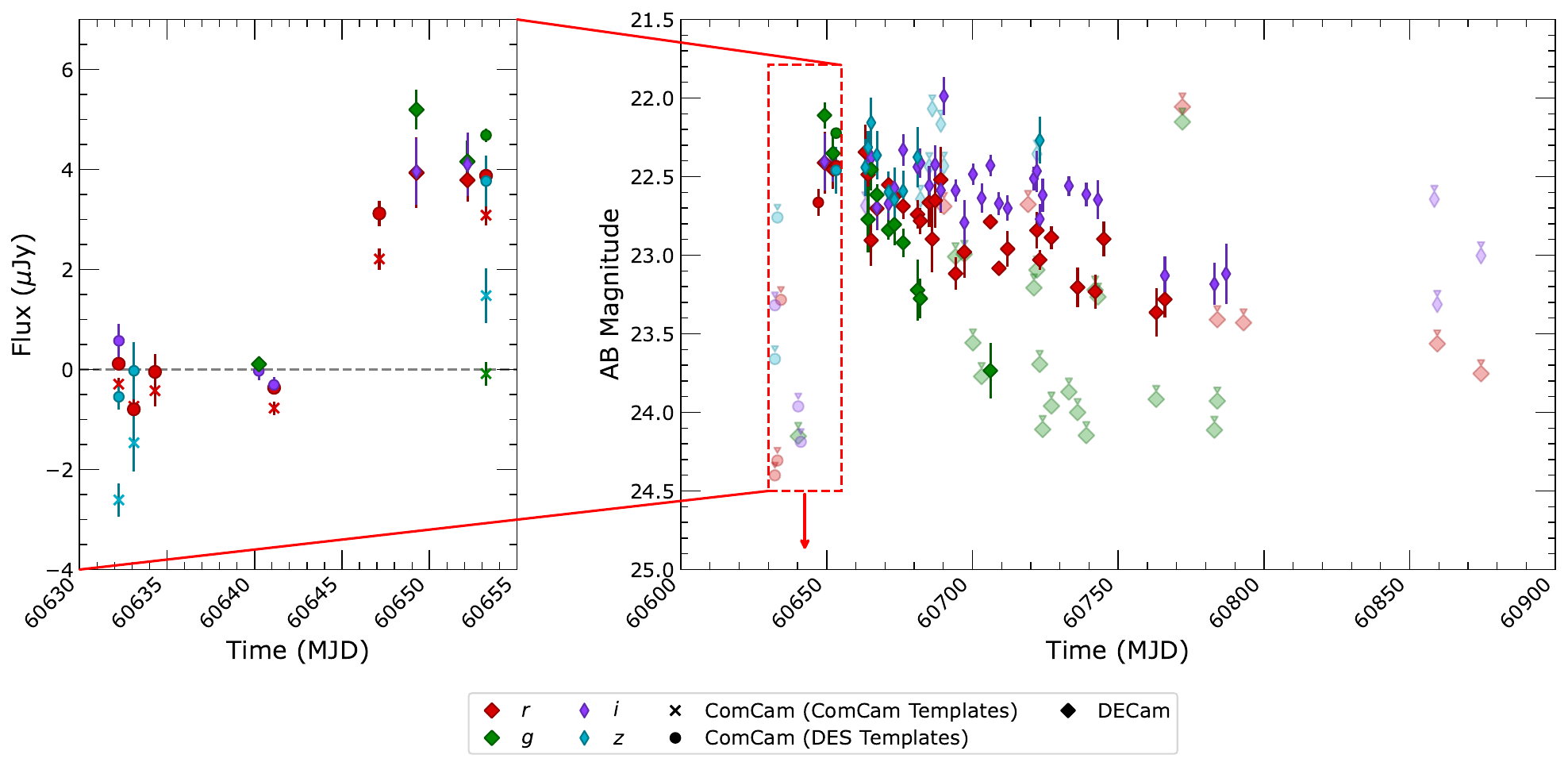}
    \vspace{1em}
    \caption{\textit{Left:} Comparison of ComCam $griz$-band difference-image photometry of AT~2024ahzi as queried directly from RSP (i.e., using the ComCam templates; circles), versus corrected using robust DES templates (squares). The DECam difference-image photometry at early times is also included (triangles) to highlight the improved continuity between instruments when using the DES-subtracted ComCam photometry. \textit{Right:} The combined AT~2024ahzi light curves in magnitude space. Translucent points represent $3$-sigma upper limits of non-detections (SNR $\leq 3$). For clarity, only the DES-subtracted ComCam photometry is shown.}
    \label{fig:template_correction}
\end{figure*}

DP1 provides forced photometry extracted from ComCam difference images. However, the templates used for difference imaging are a coadded subset of the ComCam science images and contain the SN signal itself. We therefore use deep coadded DECam images from the Dark Energy Survey Data Release 2 (DES DR2, \citealt{DES2016MNRAS.460.1270D,Abbott2021ApJS..255...20A}) as templates for difference imaging. These DECam images were taken between 2013 and 2019, well before the SN explosion. The image subtraction and photometry extraction are performed using the \texttt{SLIDE}\footnote{\url{https://github.com/yizedong/SLIDE}} package \citep{Dong2025arXiv250722156D}, which is designed for LSST image subtraction using DECam templates and can run directly on the Rubin Science Platform. We only apply this template subtraction on $griz$ ComCam images, since there are no appropriate $u$- and $y$-band DES templates to use.

Individual observations within the same night and with the same filter are combined using a weighted average. As the provided uncertainties are generally much smaller than the spread among same-night observations, we instead quantify each nightly flux uncertainty as the median absolute deviation of per-night fluxes in each filter.

We also use \texttt{SLIDE} and DES templates to re-subtract DECam images, rather than the \textit{Photpipe} pipeline detailed above. We do this because DES's images are deeper than DECam's, resulting in less noisy subtractions for extremely faint observations. This only has a significant impact on the $g$-band observations, which are $\mathrm{SNR} < 3$ non-detections along the plateau; as we are working in flux space, we can still use these observations to constrain plateau properties.

Because DECam and ComCam have similar filter profiles, we combine observations across the two instruments. DECam's $z$ bandpass has a secondary bump around 9700 angstroms that ComCam lacks. We argue, however, that since there is only one ComCam $z$-band observation, the filter mismatch has a minimal impact on downstream analysis. The resulting nightly co-added $griz$-band photometry after improved template subtraction is in Table~\ref{tab:phottab}. We highlight the improvement in difference-image ComCam photometry compared to the DP1 values in Figure~\ref{fig:template_correction}. Improved continuity from ComCam to DECam photometry with the corrected templates is especially visible in the corrected $g-$band data. From MJD 60652 to MJD 60653, there is an $8\sigma$ $g$-band jump between the DECam and ComCam obesrvations using ComCam templates; with DES templates, the two observations agree within $1\sigma$.

AT~2024ahzi's first detection is $22.66 \pm 0.09$ mag, taken by ComCam at MJD 60647 in the $r$-band. To confirm that this is a real detection, we apply the \texttt{DETECT} pipeline \citep{geron2024detect} to estimate the detection thresholds for the corresponding template and visit images. We find 50\% and 80\% detection thresholds of $>23.52$ mag and $>23.49$ mag, respectively. Our observation is brighter than these thresholds and, therefore, considered a true detection. We constrain AT~2024ahzi's time of explosion between December 3, 2024 (MJD 60641) and December 9, 2024 (MJD 60647).

\section{Host Galaxy Analysis}\label{sec:host}

\begin{figure*}[t]
    \includegraphics[width=0.8\textwidth]
    {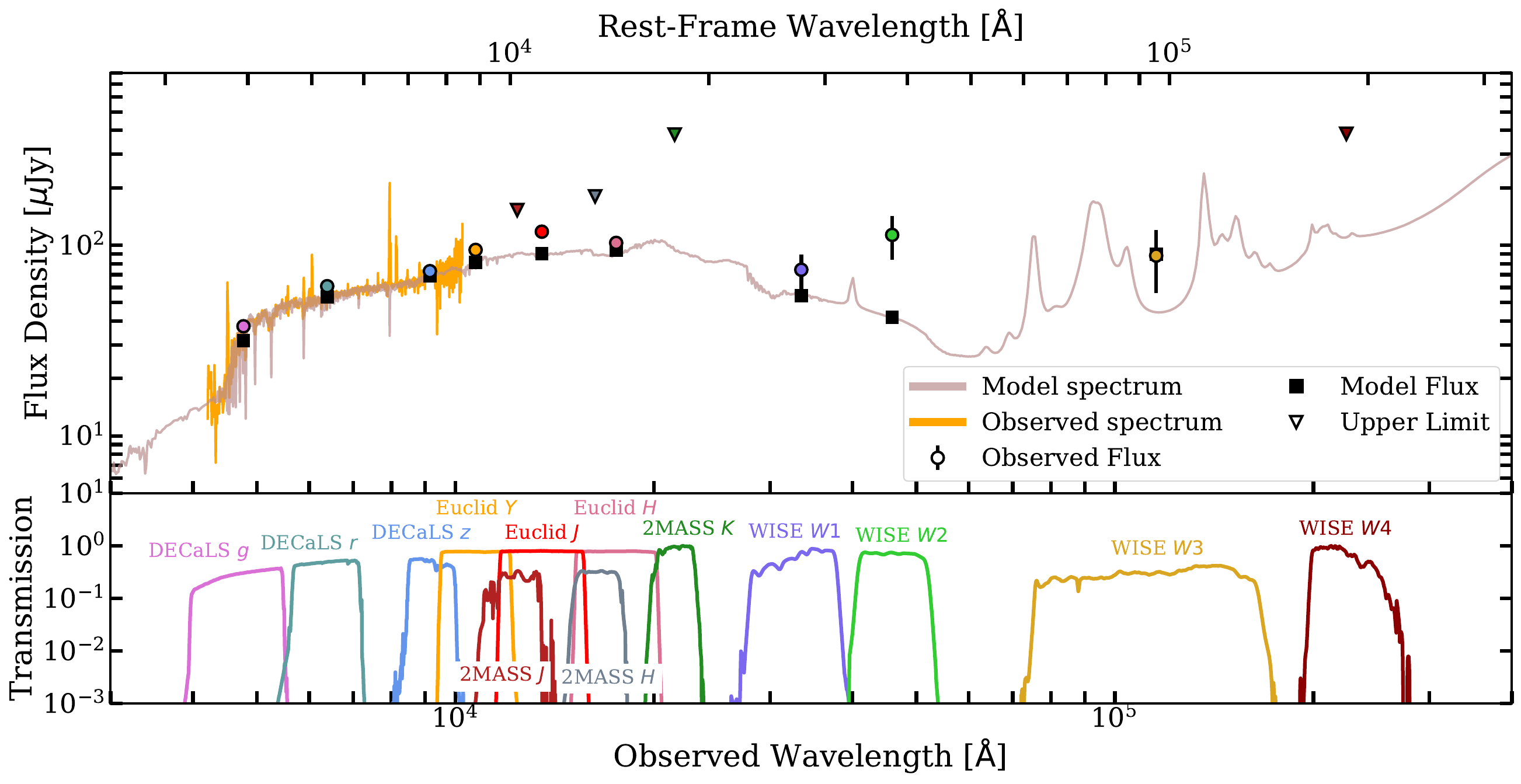}
    \caption{Observed photometry (colored circles) and spectrum (gold line) obtained of AT~2024ahzi's host galaxy, compared to the \texttt{Prospector} model spectrum (purple line) and corresponding photometry (black squares). The model suggests that the host is actively star-forming, consistent with the blue observed color of the host in the composite image.}
    \label{fig:host_sed}
\end{figure*}

Because both redshift and dust extinction can significantly alter AT~2024ahzi's photometry and therefore our downstream analysis, we first constrain these values through host galaxy association.

In Figure~\ref{fig:host} AT~2024ahzi is coincident with WISEA J035320.41-484500.0, a face-on spiral galaxy located at R.A. = $03^{\text h} 53 ^{\text m} 20^{\text s}.304$, Dec = $-48^{\circ} 44' 59^{''}.28$. We formally associate the host galaxy of AT~2024ahzi to its host galaxy using \texttt{Pr\"{o}st}\footnote{\url{https://github.com/alexandergagliano/Prost}}, a Bayesian transient association code \citep{Prost}. From the galaxy's brightness distribution and the offset between SN and galaxy, \texttt{Pr\"{o}st} confidently associates AT~2024ahzi to WISEA J035320.41-484500.0 with a chance coincidence probability of 0.002 (using the formula in \citealt{bkd02}). With a host-SN offset of 2.107'' and a host $g$-band half-light radius of 2.111'', we retrieve a relative offset $\approx 1$, which is within the interquartile range of the SN~II sample from \citep{Kelly_2012}. The host galaxy is recovered from the DECaLs DR9 catalog, with a photometric redshift of $z_\mathrm{phot}=0.202 \pm 0.058$.

The photometric redshift uncertainty corresponds to an uncertainty in distance modulus that is too large for downstream SN analysis. To extract a more precise redshift, we obtain a spectrum of the host using the Low Dispersion Survey Spectrograph (LDSS-3; \citealt{ldss}) on the Magellan Clay telescope (P.I. Gagliano). We observe the [OII]$\lambda3727$, H$\beta$, [OIII]$\lambda\lambda$4959, 5007, [NII]$\lambda\lambda$6548, 6584, H$\alpha$, and [SII]$\lambda\lambda$ 6717,6731 emission lines. Fitting these lines with a Gaussian distribution, we determine that the host lies at $z=0.211 \pm 0.002$. This is statistically consistent with the photo-$z$ recovered by \texttt{Pr\"{o}st}.

We use a customized version of \texttt{Blast} \citep{Blast}, which we call \texttt{FrankenBlast} \citep{FrankenBlast}, to extract host-galaxy photometry. \texttt{FrankenBlast} collects images of the host galaxy from DECaLS DR9, the \textit{Two Micron All-Sky Survey} (2MASS; \citealt{2MASS}), and the \textit{Wide-field Infrared Survey Explorer} (WISE; \citealt{WISE}). For every archival image, \texttt{FrankenBlast} constructs elliptical apertures surrounding the host with the \texttt{Astropy photutils} package \citep{photutils}. These apertures are optimized for each image independently. In the 2MASS \textit{JHK} and WISE \textit{W4} bands, where the host is not detected, we use the aperture size of a neighboring filter. This aperture optimization differs from the method used in \cite{Blast}, which uses consistent, PSF-adjusted aperture sizes across all wavelengths. As a result, \texttt{FrankenBlast} extracted fluxes are less noisy, but potentially underestimated in the IR and UV.

To complement the photometry retrieved with \texttt{FrankenBlast}, we collect \textit{YJH} photometry from \textit{Euclid} \citep{Euclid}. \textit{Euclid} photometry is retrieved from the \textit{Euclid} Q1 MER final catalog \citep{Romelli2025} using the \texttt{astroquery} package \citep{astroquery}. NIR fluxes are extracted from single Sersic model fitting, as measured with Source Extractor \citep{1996A&AS..117..393B,Kuemmel2022}, and converted to AB magnitudes. These fluxes should be extremely similar to those extracted directly from Euclid images, and therefore do not bias Prospector's SED fitting.

All galaxy images used for photometry extraction were taken well before AT~2024ahzi's explosion, so there is no risk of flux contamination from the SN itself. We list host magnitudes and upper limits from all sources in Table~\ref{tab:host_phot}.

\begin{table}
    \centering
    \begin{tabular}{l|c|c}
    \hline
    Survey & Filter & AB Mag \\ \hline
    DECaLS & $g$ &  19.96$\pm$0.01 \\
    DECaLS & $r$ & 19.43$\pm$ 0.01\\
    DECaLS & $z$ & 19.24$\pm$0.02  \\
    2MASS & $J$ & $>$18.44 \\
    2MASS & $H$ &  $>$18.26 \\
        2MASS & $K$ & $>$17.45 \\
         \textit{Euclid} & $Y$ &  18.96$\pm$0.01\\
         \textit{Euclid} & $J$ &  18.87$\pm$0.01 \\
         \textit{Euclid} & $H$ & 18.72$\pm$0.01 \\
         WISE & \textit{W1} & 19.22$\pm$0.24 \\
         WISE & \textit{W2} & 18.76$\pm$0.32 \\
         WISE & \textit{W3} &  19.03$\pm$0.48 \\
         WISE & \textit{W4} & $>$17.44 \\
    \end{tabular}
    \caption{Photometry of the host galaxy of AT~2024ahzi. All photometry is corrected for dust extinction in the direction of the SN, using the \citet{sf11} dust maps and $A_V=0.30^{+0.08}_{-0.07}$.}
    \label{tab:host_phot}
\end{table}

To model the stellar population properties of the host, we jointly fit the photometric detections and spectrum with \texttt{Prospector} \citep{Leja2019, jlc+2021}, a \texttt{Python}-based stellar population modeling inference code. Sampling and model details can be found in Appendix~\ref{sec:prospector}. The SED of the best-fit model is shown in Figure~\ref{fig:host_sed}. We find that the host of AT~2024ahzi has a stellar mass of log($M_*/M_\odot$)=$9.86^{+0.03}_{-0.03}$, which is consistent with previous samples of both low-redshift \citep{Schulze_2021} and high-redshift \citep{Grayling_2023} SN~II host galaxies. It also has a star formation rate of $1.38^{+0.886}_{-0.50} M_\odot$~yr$^{-1}$ and specific SFR of $\log(\mathrm{sSFR})=-9.8^{+0.10}_{-0.09}$~yr$^{-1}$. These values are typical for SN II host galaxies, as shown in \citep{Schulze_2021}. We determine that the host lies along the star-forming ``main sequence'' as defined in \cite{Leja_2022} and \cite{Popesso_2022}, and is therefore actively star-forming. This is consistent with our classification of AT~2024ahzi as a SN~IIP; a quiescent host galaxy would be far less likely to house a core-collapse SN. 

\texttt{Prospector} returns a $V$-band dust attenuation of $A_V$=$0.30^{+0.08}_{-0.07}$~mag, but the effective reddening of AT~2024ahzi's photometry depends on both the host's offset and line of sight. Therefore, in the next sections we analyze AT~2024ahzi's photometry without correcting for host extinction unless otherwise stated.

\section{Photometric Analysis} \label{sec:analysis}

In this section, we discuss all processing and downstream analysis of AT~2024ahzi's DECam and ComCam photometry.

There are 187 $griz$-band observations of AT~2024ahzi in the combined ComCam-DECam photometry. We convert the photometry from apparent to absolute magnitudes using $z_{\mathrm{host}} = 0.202$ and a flat $\Lambda$CDM cosmology consistent with the Planck 2015 results \citep{planck_2016}. We also phase the light curve around peak, and correct the time axis for time dilation. We treat ComCam and DECam observations in the same passband as from a single instrument; this is a fair assumption since they share similar filter profiles and overlap in their photometry (as shown in Figure~\ref{fig:template_correction}).

Additionally, we correct for Milky Way extinction using the dust maps provided by \cite{dustmaps1} and \cite{dustmaps2}, and assuming a \cite{fitzpatrick_massa_2007} extinction model with $R_V=3.1$. We repeat the following analysis with and without a correction for host extinction, and compare the two sets of results.

\subsection{Type IIP Classification and Superphot+ Fitting}\label{subsec:classification}

\begin{figure*}[t]
  \centering
  
  \begin{subfigure}{0.5\textwidth}
      \centering
      \includegraphics[width=\textwidth]{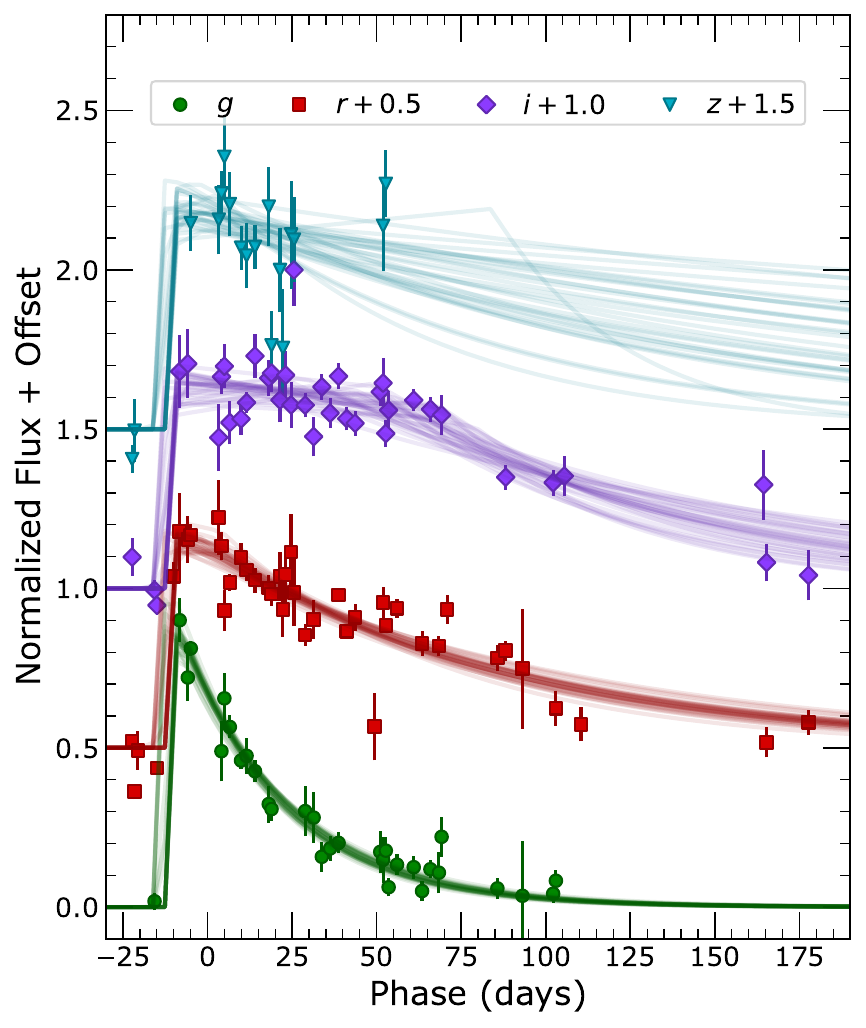}
    \end{subfigure}%
    \begin{subfigure}{.5\textwidth}
      \centering
    \includegraphics[width=\textwidth]{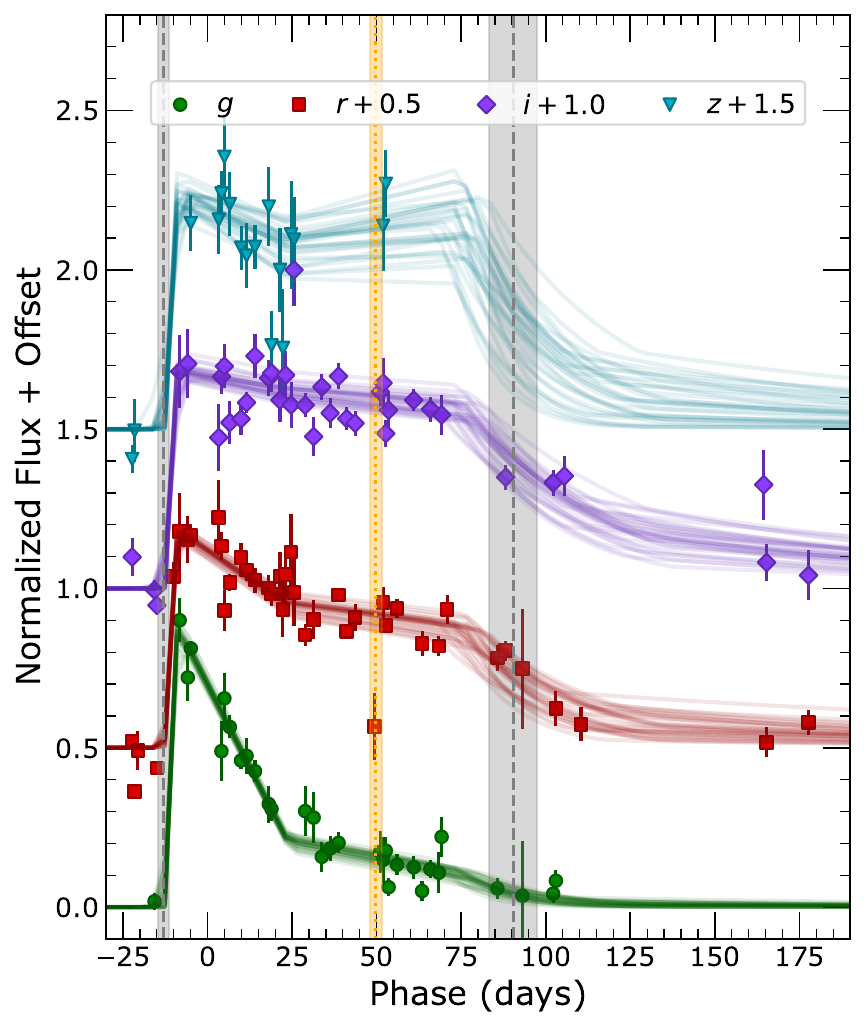}
    \end{subfigure}
  \caption{\textit{Left:} Superphot+ fit to AT~2024ahzi used for initial classification. Both the redshift-inclusive and redshift-agnostic classifiers assign a SN II label to the best-fit parameter set, with confidences of 98\% and 92\%, respectively. \textit{Right:} SN~IIP-specific Superphot+ fit to AT~2024ahzi. The flat plateau from hydrogen recombination is differentiated from the steeper dimming and cooling region following shock-breakout, possibly lengthened by an extended CSM. We define ``plateau duration'' as the time from explosion to where the flux drops halfway after hydrogen recombination, as bounded by the vertical gray dashed lines. We measure the plateau magnitude and color from midway along the hydrogen recombination region (dotted yellow line)}
  \label{fig:sp_fit}
\end{figure*}

Visually, the fast rise and extended plateau of AT~2024ahzi is photometrically consistent with a SN~IIP. As a more quantitative measure, we use Superphot+~\citep{superphotplus} to both fit and classify the light curve of AT~2024ahzi. Superphot+ fits multiband photometry to user-defined parametric models using a suite of samplers and utilizes these parameterized fits for photometric classification.

We fit our light curves jointly across filters (normalized in flux space) using a standard empirical SN model (\citealt{Villar_2019}, \citealt{superphot}), as shown in the left panel of Figure~\ref{fig:sp_fit}. Using the best-fit model values as inputs, we classify AT~2024ahzi using two supernova classifiers from \cite{superphotplus} that include and exclude redshift information, respectively. These classifiers yield one of five output labels: SN~Ia, SN~II (which includes SNe~IIP and SNe~IIL), SN~IIn, SN~Ib/c, and SLSN-I. Because the ZTF public surveys only observe in the $g$- and $r$-bands, we use our fit's $g$- and $r$-band parameters for classification. The redshift-agnostic ZTF classifier reports a 92\% SN II pseudo-probability, while including peak absolute magnitude as a feature raises the pseudo-probability to 98\%. We therefore photometrically classify AT~2024ahzi as a SN II with high confidence. We further subclassify it as a SN IIP from its visually extended plateau, though recent analysis supports a continuum between SNe~IIP and SNe~IIL over a clear distinction \citep{snii_continuum, snii_continuum2, Morozova_2017, Hiramatsu_2021, Anderson_2014, Sanders_2015, valenti2016, Galbany_2016}.

We note that this is an imperfect classification method, since the classifiers are trained on data from ZTF, and there is a filter mismatch between the ZTF and LSST filters. Furthermore, we do not apply a $K$-correction to account for the wavelength shift between the emitted and observed flux distribution. The latter effect is significant, since the ZTF training set is also low redshift ($z \lesssim 0.1$). To highlight the robustness of our classification to $K$-corrections, we use AT~2024ahzi's $r$- and $i$-band fit parameters as inputs, mimicking the full filter shift in emitted $gr$-band flux that would occur if $z \approx 0.22-0.28$. The classification pseudo-probability decreased to 94\% and 87\% with and without redshift information, respectively. If we adjust the original $g$- and $r$-band fit parameters by correcting for host extinction, the pseudo-probabilities are 97\% and 95\%. Because Superphot+ SN II pseudo-probabilities are underconfident, all of the above pseudo-probabilities correspond to calibrated confidences $>99\%$. Therefore, we conclude that our SN~II classification for AT~2024ahzi is robust to both redshift effects and host-galaxy dust.

In practice, we want to classify Rubin transients as early as possible for timely spectroscopic follow-up. Even when only ComCam (i.e., early-phase) photometry is used for fitting and classification, Superphot+ still labels AT~2024ahzi as a SN~II, albeit with lower confidence. A modified classifier trained only on early-phase fit features (see Section 5.2 of \citealt{superphotplus}) returns a SN~II prediction with 63\% pseudo-probability (90\% calibrated confidence) with redshift information and 53\% pseudo-probability (84\% calibrated confidence) without redshift information. This suggests that SNe~II can be discerned at very early phases from their fast rise timescales \citep{gonzalezgaitan2015, Gal_Yam_2022}.

After using the general SN model fit to classify AT~2024ahzi, we refit the photometry to a parametric model specifically designed for SNe~IIP, fitting a steeper, shock cooling component preceding the shallower recombination plateau \citep{de_soto_in_prep}:

\begin{align}
F(\phi) = &\;\dfrac{A}{1 + e^{-\phi/\tau_\mathrm{rise}}} \times \nonumber \\
&\begin{cases}
1 - (\beta_1 + \beta_2)\phi, & \text{if } \phi < \gamma_1 \\
1 - \beta_1 \gamma_1 - \beta_2\phi, & \text{if } \gamma_1 \leq\phi < \gamma_2 \\
C\exp\left[\dfrac{\gamma_2 - \phi}{\tau_\mathrm{fall}}\right], & \text{if } \gamma_2 \leq \phi < \gamma_\mathrm{cobalt} \\
\begin{array}[t]{@{}l@{}}
C\exp\left[\dfrac{\gamma_2 - \gamma_\mathrm{cobalt}}{\tau_\mathrm{fall}}\right] \\
\quad \times \exp\left[\dfrac{\gamma_\mathrm{cobalt} - \phi}{\tau_\mathrm{cobalt}}\right]
\end{array}, & \text{if } \phi > \gamma_\mathrm{cobalt}
\end{cases}
\end{align}

 Here, $F(\phi)$ is the normalized flux in each band, $\phi$ is a phase, and $\tau_\mathrm{rise}$, $\tau_\mathrm{fall}$, and $\tau_\mathrm{cobalt}$ are rise, fall, and radioactivate decay timescales, respectively. The $\beta$ and $\gamma$ parameters correspond to slopes and transition points of each linear decline. The first piecewise component represents the initial rise and dip, the second component corresponds to the shallower plateau region, the third component corresponds to the post-plateau brightness decline, and the last component is the cobalt decay tail. This is similar to the functional form found in \cite{superphotplus}, but now there is differentiation between a steeper post-rise decline ($\beta_1$ and $\gamma_1$), a shallower plateau ($\beta_2$ and $\gamma_2$), and a radioactive decay tail ($\beta_\mathrm{cobalt}$). This model is used to fit each band separately, but joint priors ensure that the piecewise turning points agree across filters. The best fit to the model is shown in the right panel of Figure~\ref{fig:sp_fit}. To determine whether the added SN~IIP-specific model parameters improve the fit to a statistically significant degree, we calculate the Bayesian information criterion (BIC) for both the simpler classification model and the SN~IIP-specific model. A lower BIC indicates that extra degrees of freedom (i.e. more complex models) are sufficiently improving the fit. The SN~IIP fit has a BIC of 149, while the classification SN fit has a BIC of 174. Therefore, we conclude that a parametrization which includes both a steeper and shallower linear region is needed to effectively fit AT~2024ahzi's photometry. This suggests an extended (possibly CSM) component around AT~2024ahzi's progenitor, as CSM-ejecta interaction is expected to produce early, blue emission.

From the IIP-specific fit, we extract a plateau duration of $103.4 \pm 7.3$ days. For this analysis, we define the plateau duration as the time from explosion to midway between plateau end and cobalt tail (e.g., see \citealt{valenti2016} and \citealt{goldberg_2019}), indicated by the dashed gray lines in the right panel of Figure~\ref{fig:sp_fit}. Note that this differs from both the OPTd and Pd definitions in \cite{Anderson_2014} and \cite{gutierrez-2017} but is equivalent to the `optd' defined in \cite{Martinez_2022}. The latter argues that plateau duration as defined here is more strongly correlated with hydrogen envelope mass than other definitions. 

We also extract a mid-plateau $r$-band absolute magnitude of $-17.39 \pm 0.09$ mag. We apply a $K$-correction to this magnitude after estimating the temperature evolution in Section~\ref{sec:extrabol}.

\subsection{Bolometric Evolution}\label{sec:extrabol}

\begin{figure}[t]
    \includegraphics[width=\linewidth]{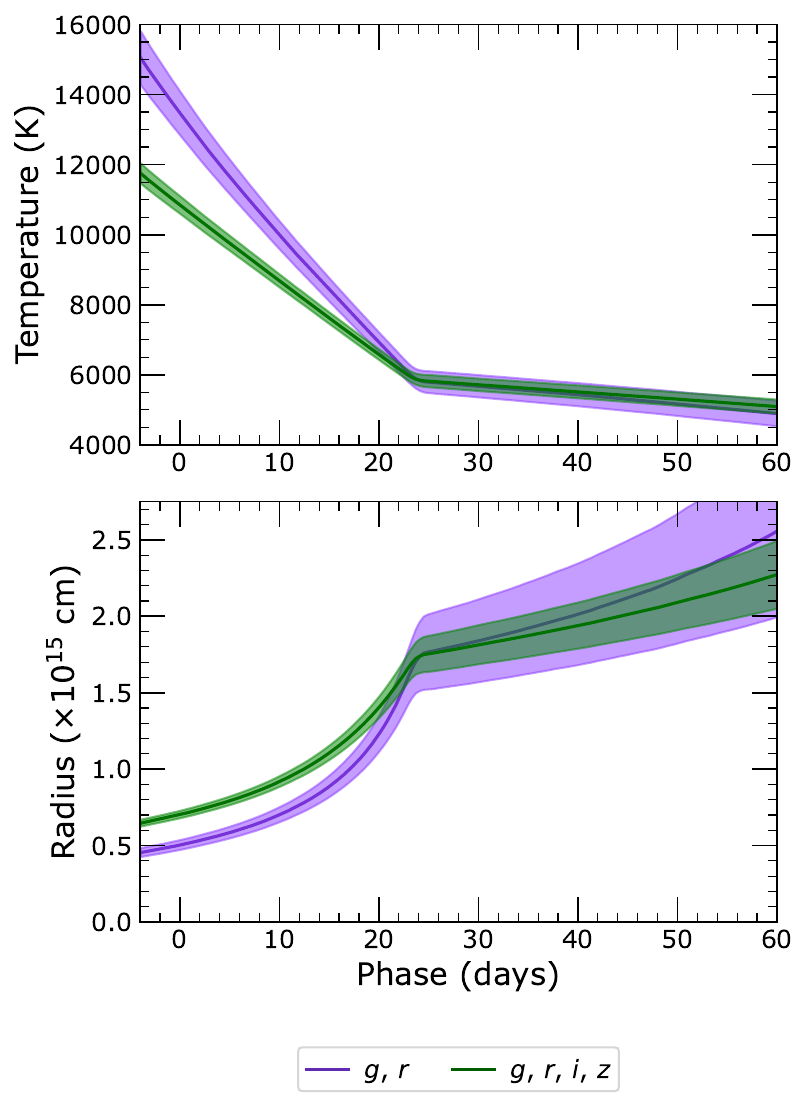}
    \caption{Blackbody evolution of AT~2024ahzi, as fit by \texttt{extrabol}, using both $griz$- and only $gr$-band data. The two fits agree along the plateau but differ at early times, a bias that persists after adjusting for host extinction.}
    \label{fig:bb_evolve}
\end{figure}

We use the IIP Superphot+ fit to interpolate AT~2024ahzi's multi-band light curve. From this interpolated photometry, we estimate AT~2024ahzi's bolometric evolution using the code \texttt{extrabol}~\citep{extrabol}, which approximates the SN~IIP photosphere during rise and plateau as an evolving blackbody (see e.g. \citealt{iip_bb_assumption, iip_bb_dilution} for caveats regarding flux dilution), and fits each epoch of its multiband photometry to a Planck function. When we jointly fit $griz$-band data, we recover an effective temperature of $T_{\mathrm{eff}} \approx 11000$ K and photospheric radius of $R_{\mathrm{phot}} \approx 5 \times 10^{14}$ cm at peak. Along the plateau, we recover $T_{\mathrm{eff}} \approx 6000\mathrm{ K}$, which matches the recombination temperature for hydrogen, and $R_{\mathrm{phot}} \approx 2 \times 10^{15}$ cm. This plateau temperature rules out the possibility of AT~2024ahzi being a SN~IIn, which is important to keep in mind when studying its early-phase evolution in Section~\ref{subsec:csm_sec}.

We also fit an evolving blackbody using only $g$- and $r$-band data, which is compared to the four-band fit in Figure~\ref{fig:bb_evolve}. The two fits align closely along the plateau, but the $gr$-fit estimates higher temperatures at early phases. This offset remains even when the photometry is corrected for host extinction. We can attribute this offset to a slight excess of $iz$-flux; since all observations lie in the blackbody tail for these temperatures, slight flux excesses can dramatically shift the extracted temperature and radius. This is a trend also seen in \cite{Faran_2018} and explained by \cite{tominaga_2011} and \cite{shussman_2016}; higher-frequency photons probe deeper and therefore hotter layers of the ejecta.

Using the effective temperature evolution derived from the \texttt{extrabol} fit, we estimate $K$-corrections assuming $z=0.211$ and a blackbody SED, propagating uncertainties in both temperature and redshift. 

\subsection{Progenitor and CSM Characterization}\label{subsec:csm_sec}

A supernova's photometry reveals information about the progenitor that exploded, as well as the CSM surrounding it at the time of explosion. Observed photometry can be cross-matched with a grid of modeled light curves to infer the progenitor, explosion, and CSM properties (e.g., \citealt{Morozova_2017,Eldridge_2019, Hiramatsu_2021,Barker_2022,Martinez_2022,moriya}). However, various progenitor configurations can yield statistically identical light curves, though many such configurations are ruled out by stellar evolution models that assume a unique mapping from progenitor radius to ejecta mass \citep{kasen_2009, dessart_2019, goldberg_2019,goldberg_2020,dessart_2023,progenitor_pulsations,Fang2025a,Fang_2025_2}. For example, the progenitor models in the \cite{moriya} grid are obtained from \citet{sukhbold_2016} and are characterized solely by one of five zero-age main sequence (ZAMS) masses ($M_{\mathrm{ZAMS}}/M_\odot\in \{10,12,14,16,18\}$), where solar metallicity is assumed, and therefore an implicit relation between pre-explosion envelope mass and progenitor radius is established.

To constrain the degeneracy in progenitor and explosion properties for AT~2024ahzi without introducing biases from a model grid, we plug its plateau duration and mid-plateau luminosity ($L_{\mathrm{plat}} = 2.48 \times 10^{42}$ erg/s, extracted from the blackbody fit) into the analytic scalings from \cite{goldberg_2019} and \cite{goldberg_2020}:
\begin{multline}
    \log L_{\mathrm{mid}} = 42.16 - 0.40 \log M_{10} + 0.74 \log E_{51} \\ + 0.76\log R_{500}
\end{multline}
\begin{multline}
    \log t_p = 2.184 + 0.134 M_{\mathrm{Ni}} + 0.411 \log M_{10} \\ - 0.282 \log E_{51}
\end{multline}

These scalings relate progenitor radius ($R_{500} = R_0 / 500 R_\odot$), H-rich ejecta mass ($M_{10} = M_{\mathrm{ej}} / 10 M_\odot$), and combined kinetic and radiative explosion energy ($E_{51} = E_{\mathrm{exp}} / 10^{51}$ erg), for a specified nickel mass $M_{\mathrm{Ni}}$. We set $M_{\mathrm{Ni}} = 0.1 M_\odot$ in accordance with the $M_{\mathrm{Ni}}-L_{50}$ relation derived in \cite{nickel_distro}, satisfying the $M_{\mathrm{Ni}} \gtrsim 0.03 M_\odot$ criterion assumed by the scalings. The two plateau properties define a curve in radius-mass-energy space rather than a single set of values. \cite{goldberg_2019} estimates a $\sim$15\% theoretical uncertainty in values extracted from this curve.

Considering the broad range of potential progenitor radii supported by \cite{prog_radius_range} of $250 R_\odot$ to $1750 R_\odot$, the possible H-rich ejecta mass and explosion energy range from $1.58 - 12.13 M_\odot$ and $(0.16 - 2.6) \times 10^{51}$ erg, respectively. Specifically, larger values for the progenitor radius correspond to lower inferred H-rich ejecta masses and explosion energies. An independent constraint on progenitor radius is needed to better constrain H-rich ejecta mass and explosion energy. Direct progenitor observation \citep{rsg_detection} or a measurement of its variability \citep{progenitor_pulsations} is only possible for nearby systems (distance $\lesssim 30$ Mpc). We can potentially constrain progenitor luminosity from nebular spectra \citep{Fang_2025, Fang2025c}, which requires spectroscopic follow-up. In the absence of direct or spectroscopic constraints, as will be the case for most Rubin transients, early ($\lesssim$15 days after explosion) estimates of photospheric velocity may constrain explosion energy in the absence of CSM interaction \citep{goldberg_2019, Dessart_2010}. For events affected by CSM, we can still constrain families of progenitor and explosion properties, which is useful for ruling out invalid stellar evolution or explosion models.

 The prevalence and diversity of CSM around SNe~IIP remain an open question \citep{yaron_2017, irani_2024, jacobson_2024, iip_csm_ztf}, but its impact on optical emission tends to be restricted to early phases ($\lesssim 40$ days; \citealt{bruch_2023}). To explore possible CSM interaction affecting AT~2024ahzi's early-phase photometry, we fit the photometry to the grid of synthetic SN~II models from \cite{moriya}. While this grid has limited progenitor diversity, the early lightcurve is mostly dependent on CSM interaction and thus can more likely be well-represented by the grid's extensive CSM parameter space. The stars are exploded using \texttt{STELLA} \citep{stella_1998}, a one-dimensional hydrodynamics code with multi-group radiative transfer. Explosions are modeled across ten explosion energies ($E_{\mathrm{exp}}/(10^{51} \textrm{ erg}) \in$ \{0.5, 1, ..., 4.5, 5\}) and nine nickel masses ($M_{\mathrm{Ni}}/M_\odot \in$ \{0.001, 0.01, 0.02, 0.04, 0.06, 0.08, 0.1, 0.2, 0.3\}), uniformly mixed to half the H envelope mass.

The CSM density profile is parameterized by an accelerating wind model, defined as:
\begin{align}
    \rho_{CSM}(r) = \frac{\dot{M}}{4\pi v_{\mathrm{wind}}(r)r^2} \\
    v_{\mathrm{wind}}(r) = v_0 + (v_\infty - v_0)\Big(1 - \frac{R_0}{r}\Big)^\beta. 
\end{align}

Here, $\beta\in \{0.5, 1,...4.5,5\}$ is a shape parameter that controls wind acceleration, $R_{\mathrm{CSM}} \in \{1,2,4,6,8,10\} \times 10^{14}$ cm is the CSM upper radius, and $\log_{10}\dot{M} \in \{-5,4.5,...-1.5,-1\}$ $M_\odot/\mathrm{yr}$ is the stellar mass loss rate; all models assume a constant mass-loss rate, and this lost mass forms the CSM. The parameters $R_0$ and $v_0$ are the progenitor radius and CSM velocity at the progenitor radius, where $v_0$ is chosen to maintain a continuous density profile at the progenitor-CSM interface. The asymptotic CSM velocity is set to $v_\infty = 10$ km/s to reflect wind velocities found around nearby RSGs \citep{van_loon_2001, goldman_2016}.

Using the SEDs from \cite{moriya}, we generate apparent multi-band Rubin light curves at $z=0.211$ to compare with the observed AT~2024ahzi photometry. To save compute, we only generate redshifted light curves for 81,842 out of the 228,016 total SEDs. We first generated light curves for a coarser distribution of parameters, then iteratively expanded the finer grid around the best-fit models until the fit quality sufficiently declined. Because the explosion time of the modeled light curve may not exactly agree with the explosion time inferred from the optical photometry, we allow for a phase dither of $\pm 20$ days to best optimize the fit. We find that many models fail to replicate both the CSM-affected (early-phase) and plateau regions. As a result, we separately fit the model grid to the $\phi < 40$ and $\phi > 40$ subsets of AT~2024ahzi's $griz$-photometry, where $\phi$ is the number of days after explosion. We call these the ``early-phase'' and ``late-phase'' fits, respectively. We only consider models with chi-squared values within 20\% of the minimum for \textit{both} fits. This ensures good agreement along the SN's entire evolution. 46 models pass this criteria, each with combined reduced chi-squared $3.2 < \chi^2_\mathrm{red} < 3.8$. The early-phase fits have $4.6 < \chi^2_{\mathrm{red}} < 5.5$, and are statistically worse than the late-phase fits, which have $2.0 < \chi^2_{\mathrm{red}} < 2.3$. Relaxing the reduced chi-squared cut to within 50\% of the minimum increases the well-fit set to 310 models. This keeps all parameter distributions approximately the same, except it extends the range of well-fit ZAMS masses.

\begin{figure}[t]
  \centering
  \includegraphics[width=\linewidth]{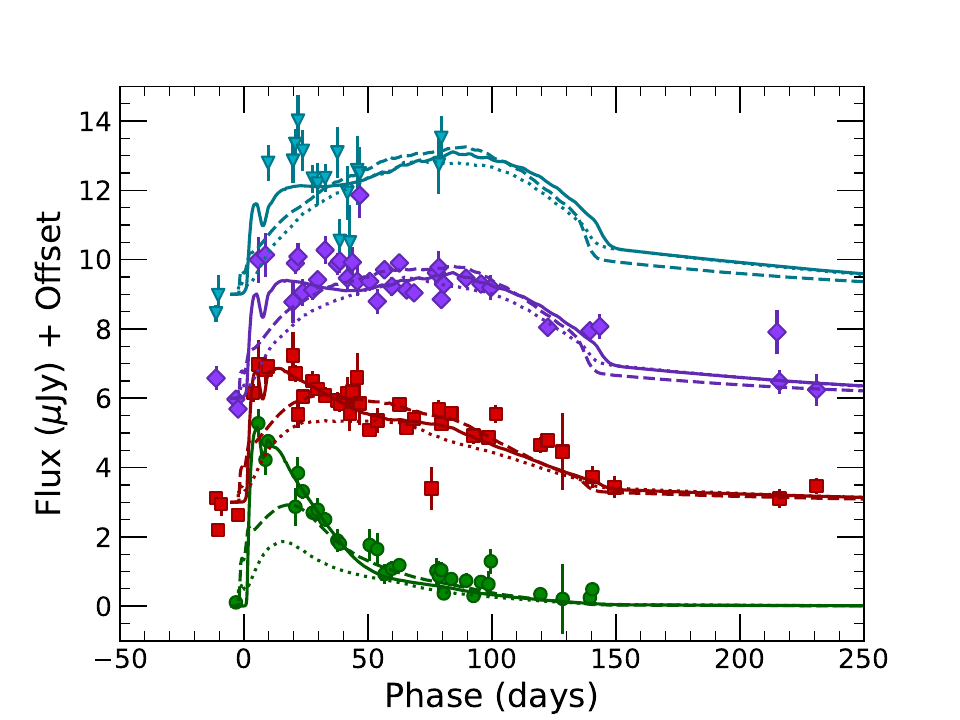}
  \caption{Optimal CSM model (solid line) from \cite{moriya}, found among the models well-fit within both the early ($\phi < 40$ days) and late ($\phi > 40$ days) phases of 2024ahzi's photometry. We compare to the best CSM-less model (dashed), and the best CSM model but with the CSM removed (dotted). Both CSM-less fits are poor relative to the best CSM fit, highlighting how CSM must be added to the \cite{sukhbold_2016} stellar profiles to replicate AT~2024ahzi's photometry. The best CSM fit favors a low mass-loss rate but slow wind acceleration, which is effectively indistinguishable from an extended stellar profile at the photosphere.}
  \label{fig:csm_fit}
\end{figure}

Within the 46 well-fit models, all have $E_{\mathrm{exp}}=10^{51}$ ergs and 96\% have $M_{ZAMS}=12-14 M_\odot$, which corresponds to $M_{\mathrm{H,ej}}=7.8-8.1 M_\odot$. The remaining have $M_{ZAMS}=10 M_\odot$ or $16 M_\odot$, and these masses become more prevalent when extending the reduced chi-squared cutoff. The fit nickel masses are between $0.06-0.1 M_\odot$, which matches the $M_\mathrm{Ni}=0.1 M_\odot$ assumption we made when deriving the \cite{goldberg_2019} scalings. The model grid intersects the radius-mass-energy curve we derived above at $M_{\mathrm{ZAMS}}$ between $12 M_\odot$ and $14 M_\odot$ and $E_{\mathrm{exp}}$ between $(1-1.5) \times 10^{51}$ ergs. Therefore, our best-fit \cite{moriya} models are consistent with the derived \cite{goldberg_2019} scalings.

The majority of well-fit models have slow wind acceleration and moderate mass loss rates; $89\%$ have $\beta \geq 2.5$ and $-2.5 \leq \log_{10}\dot{M} \leq -1.5$. This aligns with the results of \cite{silva_farfan_2024} using the \cite{moriya_2018} model grid. There is not a strong preference for $R_{\mathrm{CSM}}$, but a coefficient of determination of $R^2=0.9$ indicates that larger CSM radii strongly correlate with lower mass loss rates.

For each well-fitting model, we calculate the total CSM mass $M_{\mathrm{CSM}}$ using Equations 1 and 2 from \cite{moriya} and:
\begin{multline}
    M_{\mathrm{CSM}} = 4\pi \int_{R_0}^{R_{\mathrm{CSM}}} \rho(r) r^2 dr \\
    = \dot{M} \int_{R_0}^{R_{\mathrm{CSM}}} v_{\mathrm{wind}}^{-1}(r) dr \\
    =\dot{M} \int_{R_0}^{R_{\mathrm{CSM}}} \Big[v_0 + (v_\infty - v_0)\Big(1 - \frac{R_0}{r}\Big)^\beta\Big]^{-1} dr \\
    \label{csm_eq}
\end{multline}

Before constraining AT~2024ahzi's CSM mass, we first argue that AT~2024ahzi's early-phase photometry cannot be well-fit without an extended, CSM-like density profile. To show this, we consider only models with $M_{\mathrm{CSM}} < 10^{-2.5} M_\odot$ as a proxy for CSM-less models, in accordance with \cite{iip_csm_ztf}. The best CSM-less model has higher progenitor mass (ZAMS$=16 M_\odot$) than the best CSM model, with $\chi^2_{\mathrm{red}}=12.3$ for $\phi < 40$ days and $\chi^2_{\mathrm{red}}=7.3$ for the full light curve. This translates to a BIC of 1020 for the full light curve; when we add CSM, the lowest BIC is 475. Therefore, we conclude that the CSM parameters are needed to adequately model AT~2024ahzi's photometry, at least within the model grid used.

The best-fit CSM model, which minimizes the combined chi-squared value within our ``well-fit'' set, has $M_{\mathrm{ZAMS}} = 12 M_\odot$, $E_{\mathrm{exp}}=10^{51}$ ergs, $M_{\mathrm{Ni}}=0.1M_\odot$, $\beta=4.5$, $\dot{M} = 10^{-2} M_\odot/\mathrm{yr}$, $R_{\mathrm{CSM}}=4\times 10^{14}$ cm, and $M_{\mathrm{CSM}}=1.4 M_\odot$. The corresponding model fit is shown in Figure~\ref{fig:csm_fit}.

For all well-fit models, $M_{\mathrm{CSM}}$ ranges from 0.7 to 2.1 solar masses. These are values typical of SNe~IIn rather than SNe~II \citep{Ransome_2025}, and is on the very upper end of the CSM mass derived from \cite{iip_csm_ztf}. Because AT~2024ahzi's effective temperature remains near 6000K for the entirety of the plateau region, it is unlikely to be a misclassified SN~IIn. We conclude that either (1) CSM mass estimates are highly sensitive to the assumed CSM density prescription, and we have made assumptions different to previous works
or (2) we are overestimating the mass by assuming all is detached from the progenitor.
To support the former, we note that since we assume an accelerating wind, the degree of that wind acceleration dramatically affects the estimated CSM mass using Equation~\eqref{csm_eq}. In contrast, \cite{iip_csm_ztf} calculates $M_\mathrm{CSM}$ assuming constant wind velocity (i.e. assuming $\beta=0$ and $v_\mathrm{wind} = v(R_\mathrm{CSM})$):

\begin{align}
    M_{\mathrm{CSM,\textrm{approx}}} = \frac{\dot{M} R_\mathrm{CSM}}{v(R_\mathrm{CSM})} \\
    = \frac{\dot{M} R_\mathrm{CSM}}{v_0 + (v_\infty - v_0)(1 - R_0/R_\mathrm{CSM})^\beta}
\end{align}. Plugging our CSM values into this equation yields CSM masses between 0.1 and 0.8 $M_\odot$, which is far less anomalous compared to that paper's distribution. This highlights how large $\beta$ values can increase total CSM mass by a factor of ten by significantly slowing down the CSM wind and increasing its density.

Alternatively, we may be overestimating AT~2024ahzi's CSM mass if we are instead probing an extended, static progenitor envelope closely resembling a slowly-accelerating CSM \citep{Dessart2017}. Because the \cite{sukhbold_2016} progenitors all have an exponential density profile, a massive ``CSM'' density profile may be mimicking a greater underlying diversity in progenitor envelopes. To support this hypothesis, we only consider the CSM mass above the photosphere at time of explosion, as is the convention in \cite{davies_2022}. We recover a photosphere radius of $9 \times 10^{13}$ cm ($\approx 1300 R_\odot$), which is double the stated progenitor radius, and a total mass of $0.4 M_\odot$ above the photosphere. Additionally, the photospheric wind velocity is only 0.6 km/s, far lower than narrow line velocities observed in early phases of SNe~II with CSM interaction \citep{final_moments_i}. These negligible velocities could instead be interpreted as a more extended hydrostatic envelope, which transitions to a true wind much farther out.

While previously large RSGs have been observed in the past, this radius seems incompatible with a $12 M_\odot$ ZAMS evolutionary track \citep{Levesque_2005}. Furthermore, modeled RSG radii are likely already larger than physically viable \citep{Dessart_2013}, so a radius that further exceeds the modeled value seems unlikely.

We can reconcile our findings with these previous works by extending the progenitor model grid and relaxing the assumed relation between ejecta mass and progenitor radius in the \cite{sukhbold_2016} models. From the \cite{goldberg_2019} relations, a lower ejecta mass, smaller radius, and higher explosion energy should leave the plateau duration and luminosity unchanged. Following the light-CSM peak luminosity scalings of \cite{khatami_2024}, a higher explosion energy lowers the CSM mass needed to replicate the early-phase luminosity peak. Therefore, revised models with smaller progenitor radii may simultaneously better align with previous findings and lower estimated CSM masses to more closely match those of other SNe~IIP.

\section{AT~2024ahzi in Context}\label{sec:pop_compare}

\begin{figure}[t]
    \centering
    \includegraphics[width=\linewidth]{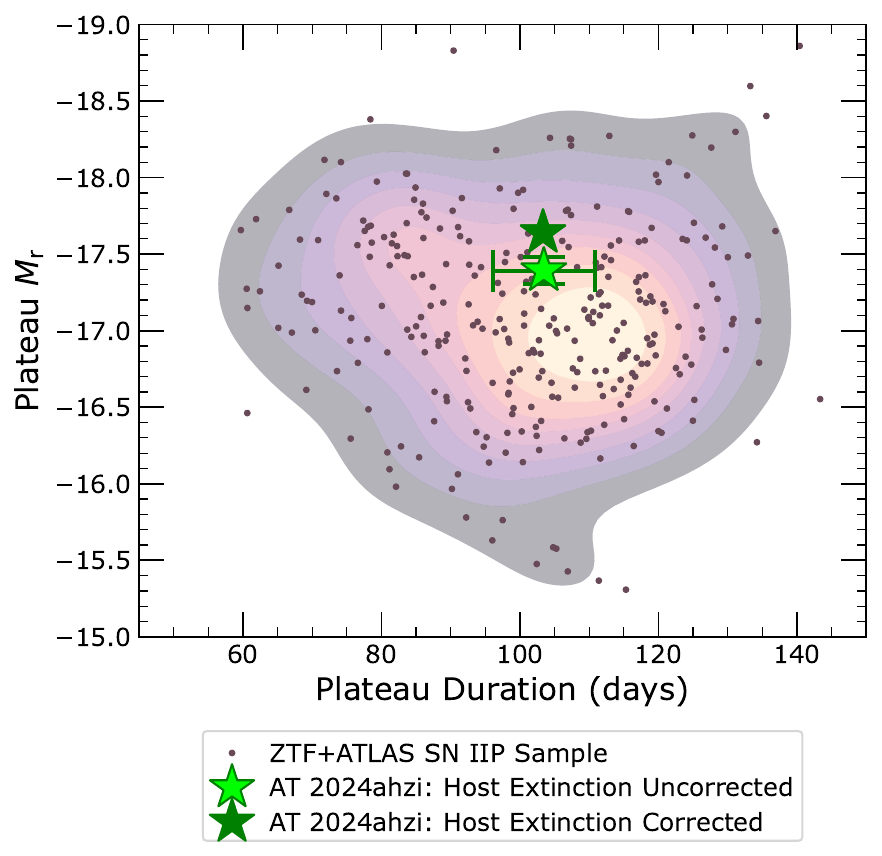}
    \includegraphics[width=\linewidth]{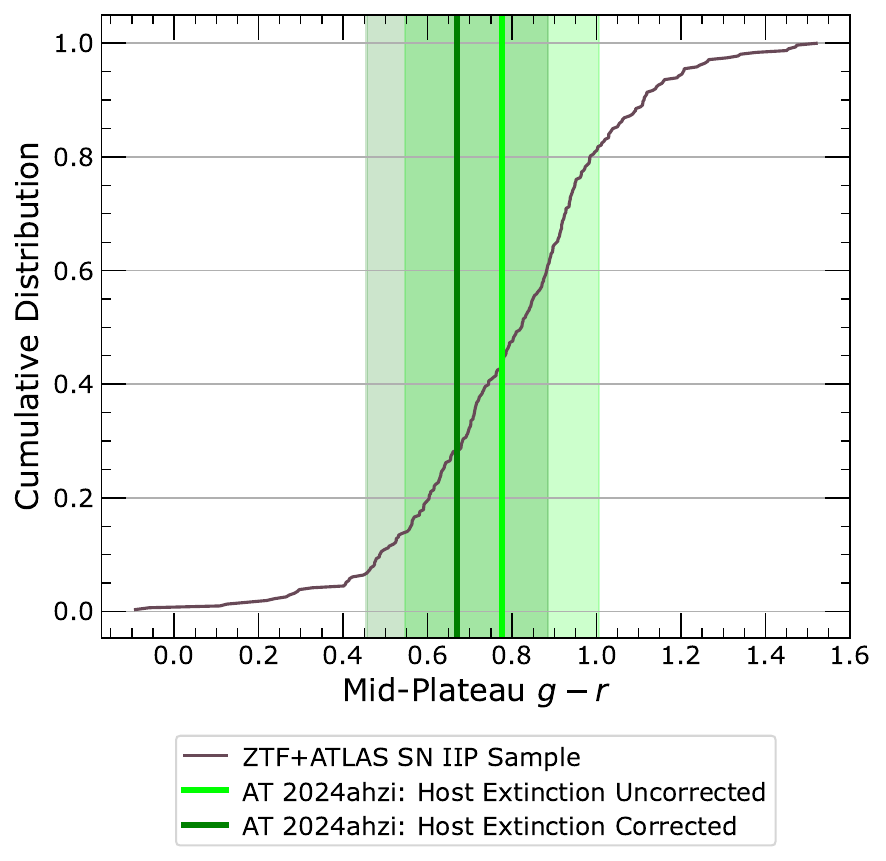}
    \caption{\textit{Top: }AT~2024ahzi's plateau duration and $r$-band absolute magnitude, as fit by Superphot+, compared to those of the ZTF-ATLAS SN~IIP sample. A blackbody $K$-correction is used to adjust AT~2024ahzi's plateau. Both correcting (dark green) and not correcting (lime green) for host extinction yields values well within the norm for the SN~IIP sample. \textit{Bottom: }2024ahzi's mid-plateau, $K$-corrected, $g-r$ color compared to the sample's cumulative distribution, before (dark green) and after (light green) correcting for host-galaxy extinction. Both $g-r$ colors are near the median for the sample.}
    \label{fig:plat_pop}
\end{figure}

We compare the plateau properties of AT~2024ahzi to a broader sample of SNe~IIP presented in \citet{de_soto_in_prep}. The sample consists of 313 spectroscopically identified SNe~IIP with high-quality photometric coverage from ZTF and ATLAS, with likely non-SNe~IIP removed via photometric cuts. The sample is low-redshift, with 90\% of the events having $z < 0.05$ and every event being closer than AT~2024ahzi. The ZTF-ATLAS sample is fit hierarchically to the SN~IIP Superphot+ model presented in Sec.~\ref{subsec:classification}.

In the top panel of Figure~\ref{fig:plat_pop}, we plot the rest-frame (e.g. corrected for time dilation) plateau duration and mid-plateau $r$-band magnitude of AT~2024ahzi against those of the larger ZTF-ATLAS sample. In the bottom panel, we compare mid-plateau $g-r$ colors. This is not a perfect comparison, however, as the Rubin $g$ and $r$ filter profiles are somewhat wider and slightly closer together than those from Palomar. All three of AT~2024ahzi's plateau properties are within 25\% of the median from the ZTF-ATLAS sample (i.e. within the interquartile range). Despite its higher redshift, AT~2024ahzi is clearly an observationally typical Type IIP SN.

Correcting for the $A_V \approx 0.30$~mag host extinction from Section~\ref{sec:host} moves AT~2024ahzi's plateau properties farther from the center of the larger SN~IIP distribution. The plateau $r$-band absolute magnitude is brighter than 66\% of the ZTF-ATLAS sample without host extinction correction, and 81\% after host extinction correction. Similarly, the $g-r$ color goes from being bluer than 57\% of the sample to 72\% of the sample after correcting for host extinction. This correction could be an overestimate, since AT~2024ahzi is located on the outskirts of its host galaxy. We note high uncertainties in AT~2024ahzi's color estimates, propagated from uncertainties in both photometric redshift and $g$-band plateau magnitude.

We do not compare AT~2024ahzi's blackbody evolution to that of the ZTF-ATLAS sample, as \texttt{extrabol}'s predictions with four-band differs from its prediction using only $gr$-band data, skewing comparison with $gr$-band ZTF-ATLAS fits.

We expect LSST SNe~II to have redshifts similar to that of AT~2024ahzi \citep{Kessler_2019}, so the LSST SN~IIP sample will be of consistently higher redshift than the ZTF-ATLAS sample. This means that the context surrounding AT~2024ahzi may change as we gather a new comparison dataset. We therefore plan on applying a similar analysis to future LSST SNe~II to quantify shifts in plateau property distributions with redshift, if any.

\section{Conclusions and Summary}  \label{sec:conc}

In this paper we present joint Rubin ComCam and DECam photometry for AT~2024ahzi. We associate AT~2024ahzi to a nearby star-forming host, which allows us to estimate the SN's redshift and dust extinction. We interpolate and photometrically classify AT~2024ahzi as a SN~IIP with $>99\%$ calibrated confidence using Superphot+, making it the first photometrically classified SN~II observed by Rubin ComCam. Because we use deep DES template images to re-extract forced photometry from ComCam images, we can strongly constrain both the brightening timescale and early color evolution of the AT~2024ahzi's light curve. This results in a confident multi-band fit and classification, even using only early-phase ComCam data. By analyzing AT~2024ahzi's multiband and blackbody evolution and comparing it to a larger population of low-redshift SNe~IIP, we determine that its properties are typical of previously observed, lower-redshift SN~IIP. Applying a similar analysis pipeline to the larger expected sample of LSST SNe~IIP will allow us to establish trends between SN~IIP light curve and progenitor properties, and quantify associated scatter with higher precision. Rather than extracting a single optimal progenitor mass and radius, we constrain a family of possible progenitor properties from the extracted plateau properties. Population progenitor analysis from many SNe~II will help us rule out models with radius-mass-energy combinations outside of the joint degeneracy curve. Finally, we fit AT~2024ahzi's early-phase photometry to a grid of CSM models and estimate a CSM mass of $0.4 M_\odot$ above the photosphere. Our fits favor slow wind acceleration in the CSM, which can alternately be explained by an extended progenitor density profile. 

Rubin is expected to discover approximately one million new transients per year \citep{tyson_2002,Kessler_2019,Graham2024_DMTN102}. Through this analysis, we outline a pipeline for identifying, processing, and studying potential SNe~II within this dataset for subsequent population studies. Because we probabilistically associate events with host galaxies, we can photometrically estimate redshift and therefore use absolute magnitude information in classification without spending limited spectroscopic resources. Our resulting dataset of photometrically classified SNe~II will elucidate the significance of the red supergiant problem \citep{smartt_2009, smartt_2015} and the prevalence of eruptive or wind-like CSM \citep{Morozova_2018, davies_2022, Matsumoto_2022}. Early-phase classification and modeling of SNe~II will allow us to rapidly follow up on those with evidence of CSM interaction; we aim to capture CSM flash features in the resulting pre-peak spectroscopy \citep{kochanek_2018, wynn_fm1, dessart_2023}.

This work highlights how observing facilities beyond Rubin can supplement Rubin photometry through simultaneous observation, effectively increasing the observing cadence. This supplementary data becomes necessary for classification when Rubin does not capture an event's full light curve evolution. Analysis of AT~2024ahzi's plateau properties is only possible with both ComCam and DECam photometry. Surveys from DECam, ZTF, YSE (through the Pan-STARRS telescopes; \citealt{panstarrs}), the \textit{Nancy Grace Roman Space Telescope} High Latitude Time Domain Survey (HLTDS; \citealt{romanhighlat}), and the La Silla Schmidt Southern Survey (LS4; \citealt{ls4}), among others, will maximize the science gleaned from transients discovered by Rubin.

\begin{acknowledgments}

The Villar Astro Time Lab acknowledges support through the David and Lucile Packard Foundation, the Research Corporation for Scientific Advancement (through a Cottrell Fellowship), the National Science Foundation under AST-2433718, AST-2407922 and AST-2406110, as well as an Aramont Fellowship for Emerging Science Research.
{K.dS.} and {C.T.M.} thank the LSST-DA Data Science Fellowship Program, which is funded by LSST-DA, the Brinson Foundation, the WoodNext Foundation, and the Research Corporation for Science Advancement Foundation; their participation in the program has benefited this work.
{J.A.G.} acknowledges financial support from NASA grant 23-ATP23-0070. The Flatiron Institute is supported by the Simons Foundation.
The UCSC team is supported in part by NSF grant AST--2307710 and by a fellowship from the David and Lucile Packard Foundation to R.J.F.
{T.G.} is a Canadian Rubin Fellow at the Dunlap Institute. The Dunlap Institute is funded through an endowment established by the David Dunlap family and the University of Toronto.
{L.I.} acknowledges financial support from the INAF Data Grant Program 'YES' (PI: Izzo) {\it Multi-wavelength and multi messenger analysis of relativistic supernovae}.
{D.A.C.} acknowledges funding through \textit{JWST} program grants JWST-GO-06541, JWST-GO-06585, and JWST-GO-05324.
{C.G.} and {D.A.F.} are supported by a VILLUM FONDEN Villum Experiment grant (VIL69896).
{D.A.F.} and {J.H.} acknowledge support from a research grant from VILLUM FONDEN (VIL54489).
{D.A.F.} acknowledges support from the VILLUM FONDEN Young Investigator Grant (project number 25501) and VILLUM FONDEN research grant VIL16599.
{W.B.H.} acknowledges support from the NSF Graduate Research Fellowship Program under Grant No. 2236415.
{D.O.J.} acknowledges support from NSF grants AST-2407632, AST-2429450, and AST-2510993, NASA grant 80NSSC24M0023, and HST/JWST grants HST-GO-17128.028 and JWST-GO-05324.031, awarded by the Space Telescope Science Institute (STScI), which is operated by the Association of Universities for Research in Astronomy, Inc., for NASA, under contract NAS5-26555.
{G.Nair's} involvement in this work is through membership in the Rubin SITCOM In-Kind Commissioning US/Chile-11 team to coordinate Rubin ComCam and DECam observations during commissioning. 
{G.Narayan} is funded by the DOE through the Department of Physics at the University of Illinois, Urbana-Champaign (\#13771275), and NSF CAREER grant AST-2239364, supported in-part by funding from Charles Simonyi.  {G.Narayan} also gratefully acknowledges NSF support from OAC-2311355 and AST-2432428, as well as AST-2421845 and funding from the Simons Foundation for the NSF-Simons SkAI Institute, and support from the HST Guest Observer Program through HST-GO-16764, and HST-GO-17128 (PI: R. Foley).
{G.Narayan} and {H.P.} gratefully acknowledge NSF support from AST-2206195. 
{M.V.} acknowledges funding from the Center for Astrophysical Surveys Graduate Fellowship.
{Q.W.} is supported by the Sagol Weizmann-MIT Bridge Program.
{A.R.W.} acknowledges support from the U.S. Department of Energy through the Department of Physics at the University of Illinois, Urbana-Champaign (\#13771275).
{Y.Z.} acknowledges support from MAOF grant 12641898 and visitor support from the Observatories of the Carnegie Institution for Science, Pasadena CA, where part of this work was completed.

This work is supported by the National Science Foundation under Cooperative Agreement PHY-2019786 (the NSF AI Institute for Artificial Intelligence and Fundamental Interactions).
Parts of this research were supported by the Australian Research Council Centre of Excellence for Gravitational Wave Discovery (OzGrav), through project number CE230100016.

The Young Supernova Experiment (YSE) and its research infrastructure is supported by the European Research Council under the European Union's Horizon 2020 research and innovation programme (ERC Grant Agreement 101002652, PI K.\ Mandel), the Heising-Simons Foundation (2018-0913, PI R.\ Foley; 2018-0911, PI R.\ Margutti), NASA (NNG17PX03C, PI R.\ Foley), NSF (AST--1720756, AST--1815935, PI R.\ Foley; AST--1909796, AST-1944985, PI R.\ Margutti), the David \& Lucille Packard Foundation (PI R.\ Foley), VILLUM FONDEN (project 16599, PI J.\ Hjorth), and the Center for AstroPhysical Surveys (CAPS) at the National Center for Supercomputing Applications (NCSA) and the University of Illinois Urbana-Champaign.

This project used data obtained with the Dark Energy Camera (DECam), which was constructed by the Dark Energy Survey (DES) collaboration. Funding for the DES Projects has been provided by the US Department of Energy, the U.S.\ National Science Foundation, the Ministry of Science and Education of Spain, the Science and Technology Facilities Council of the United Kingdom, the Higher Education Funding Council for England, the National Center for Supercomputing Applications at the University of Illinois at Urbana-Champaign, the Kavli Institute for Cosmological Physics at the University of Chicago, Center for Cosmology and Astro-Particle Physics at the Ohio State University, the Mitchell Institute for Fundamental Physics and Astronomy at Texas A\&M University, Financiadora de Estudos e Projetos, Fundação Carlos Chagas Filho de Amparo à Pesquisa do Estado do Rio de Janeiro, Conselho Nacional de Desenvolvimento Científico e Tecnológico and the Ministério da Ciência, Tecnologia e Inovação, the Deutsche Forschungsgemeinschaft and the Collaborating Institutions in the Dark Energy Survey.

The Collaborating Institutions are Argonne National Laboratory, the University of California at Santa Cruz, the University of Cambridge, Centro de Investigaciones Enérgeticas, Medioambientales y Tecnológicas–Madrid, the University of Chicago, University College London, the DES-Brazil Consortium, the University of Edinburgh, the Eidgenössische Technische Hochschule (ETH) Zürich, Fermi National Accelerator Laboratory, the University of Illinois at Urbana-Champaign, the Institut de Ciències de l’Espai (IEEC/CSIC), the Institut de Física d’Altes Energies, Lawrence Berkeley National Laboratory, the Ludwig-Maximilians Universität München and the associated Excellence Cluster Universe, the University of Michigan, NSF NOIRLab, the University of Nottingham, the Ohio State University, the OzDES Membership Consortium, the University of Pennsylvania, the University of Portsmouth, SLAC National Accelerator Laboratory, Stanford University, the University of Sussex, and Texas A\&M University.

Based on observations at NSF Cerro Tololo Inter-American Observatory, NSF NOIRLab (NOIRLab Prop.\ IDs 2024B-763968, 2024B-441839, and 2025A-388357; PI G. Narayan and A. Rest; 2012B-0001; PI: J.\ Frieman; 2014B-0404; PIs: D.\ Schlegel and A.\ Dey), which is managed by the Association of Universities for Research in Astronomy (AURA) under a cooperative agreement with the U.S.\ National Science Foundation.

This material is based upon work supported in part by the National Science Foundation through Cooperative Agreements AST-1258333 and AST-2241526 and Cooperative Support Agreements AST-1202910 and 2211468 managed by the Association of Universities for Research in Astronomy (AURA), and the Department of Energy under Contract No.\ DE-AC02-76SF00515 with the SLAC National Accelerator Laboratory managed by Stanford University. Additional Rubin Observatory funding comes from private donations, grants to universities, and in-kind support from LSST-DA Institutional Members.

No proprietary Rubin data is included in this work.

This publication makes use of data products from the Two Micron All Sky Survey, which is a joint project of the University of Massachusetts and the Infrared Processing and Analysis Center/California Institute of Technology, funded by the National Aeronautics and Space Administration and the National Science Foundation.

This work has made use of the Quick Release (Q1) data from the Euclid mission of the European Space Agency (ESA). \url{[https://doi.org/10.57780/esa-2853f3b|https://doi.org/10.57780/esa-2853f3b]}.

This publication makes use of data products from the Wide-field Infrared Survey Explorer, which is a joint project of the University of California, Los Angeles, and the Jet Propulsion Laboratory/California Institute of Technology, funded by the National Aeronautics and Space Administration.

\end{acknowledgments}

\facilities{DECam, VRO}

\software{astropy\citep{2013A&A...558A..33A,2018AJ....156..123A,2022ApJ...935..167A},  
          Cloudy \citep{2013RMxAA..49..137F}, 
          Source Extractor \citep{1996A&AS..117..393B},
          YSE-PZ \citep{coulter_d_a_2022_7278430,2023PASP..135f4501C},
          Superphot+\citep{superphotplus}
          }

\appendix
\section{Procedure for \texttt{Prospector} Modeling and Fitting}\label{sec:prospector}

\texttt{Prospector} determines posterior distributions for galactic properties using the \texttt{dynesty} \citep{Dynesty} nested sampling fitting routine, and produces model spectra and photometry using \texttt{FSPS} and \texttt{python-FSPS} \citep{FSPS_2009, FSPS_2010}. Internally, \texttt{Prospector} uses \texttt{MIST} models \citep{MIST} and \texttt{MILES} spectral libraries \citep{MILES}. Our \texttt{Prospector} model includes the \citet{kroupaIMF} initial mass function (IMF), a nebular emission model \citep{bdc+2017}, and the \citet{DraineandLi07} IR dust emission model. To measure dust attenuation in the host, we use the \citet{KriekandConroy13} dust emission model, which applies an offset to the \citet{calzetti2000} dust attenuation curve and determines the optical depth of light attenuated from old and young stellar light. To ensure that only realistic combinations of total mass formed in the galaxy ($M_F$) and stellar metallicity ($Z_*$) are sampled, we constrain these properties with the \citet{gcb+05} mass-metallicity relation. We further employ non-parametric star formation history (SFH) model with seven age-bins: the first two are linearly spaced from 0-30~Myr and 30-100Myr, and the last five are log-spaced until the age of the Universe at $z=0.211$. Finally, we apply several parameters to fit the host spectrum: a spectral smoothing model, which normalizes the spectrum to the host photometry, a gas-phase metallicity, which is dependent on the host's spectral line strengths, a template to marginalize over the observed nebular emission lines, a noise inflation model that ensures the spectrum is not overweighted in the fit compared to the photometry, and a pixel outlier model to marginalize over spectral noise. 

We use the best-fit SFH and $M_F$ to calculate a stellar mass ($M_*$), mass-weighted stellar population age ($t_m$), and present-day star formation rate (SFR). We further convert the optical depth of dust attenuated from old and young stellar light to a $V$-band dust extinction in magnitudes ($A_V$).

\bibliography{sample701}{}
\bibliographystyle{aasjournalv7}



\end{CJK*}
\end{document}